\documentclass[%
 reprint,
 amsmath,amssymb,
 aps,
showkeys]{revtex4-2}
\usepackage{subcaption}

\usepackage{graphicx}
\usepackage{dcolumn}
\usepackage{bm}
\usepackage{float}

\begin{document}

\preprint{APS/123-QED}

\title{Double circular dichroism high harmonic spectroscopy: An ultrafast probe for topological photocurrents}

\author{Osamah Sufyan}
\affiliation{%
Department of Mathematics, Informatics, and Technology, University of Applied Sciences Koblenz, 53424 Remagen, Germany
}%

\author{Ofer Neufeld}
\email{ofern@technion.ac.il}
\affiliation{%
Technion Israel Institute of Technology, Schulich Faculty of Chemistry, Haifa, 3200003, Israel
}%

\date{\today}

\begin{abstract}
\noindent ABSTRACT: Understanding optical responses of topological matter is a central problem for enabling optoelectronic applications based on topological physics, which is of fundamental concern for photocurrents control and spectroscopy. Currently, schemes for sensing ultrafast photocurrents and separating their bulk/surface contributions are lacking but are crucial for exploring topological ultrafast physics. We introduce here double circular dichroism (DCD) harmonic spectroscopy as an all-optical probe of ultrafast dynamics in topological materials. In this scheme, pump and probe pulses are circular with helicities that are independently controlled, yielding the circular dichroism of the circular dichroism — a time-resolved response evaluating how probe-induced dichroism depends on pump helicity. While DCD vanishes in symmetric systems, it survives in broken time-reversal symmetry materials including Chern insulators. We theoretically demonstrate this concept through simulations in a Haldane nanoflake, where a pump laser manipulates chiral current-carrying states, and intense probe pulses drive harmonic emission. We show that DCD originates from both bulk and edge-localized states, but these have opposite signs, similar magnitudes, and a different amplitude scaling. Moreover, edge-associated DCD is connected with the outermost atomic layer of exponentially-localized protected modes, and is relatively robust to bulk/edge interference. Hence, DCD could allow efficient separation of bulk/edge contributions to photocurrents. Variation of the electronic structure and laser parameters further reveals anomalies that might be useful for probing topological attributes of photocurrents and material phase in select harmonics. This is especially apparent in the oscillatory behavior of HHG DCD with respect to the electronic band structure manipulation which varies greatly across the phase transition and becomes stable and non-oscillatory in topologically-trivial phases. Further probing pump-probe delay we show that DCD oscillates on attosecond timescales, which could serve as an ultrafast probe of topological photocurrents and their associated coherence. Overall, our work introduces DCD as a potentially powerful approach for disentangling bulk/boundary photo-responses in broken-symmetry quantum and topological matter, and could also be implemented in other pump-probe spectroscopies based on photoelectrons and absorption, as well as chiral systems.
\end{abstract}

\maketitle


\section{Introduction} 
Topological phases of matter exhibit robust electronic properties that originate from global features of the band structure, which can drastically affect their optoelectronic responses \cite{HasanKane2010}. A paradigmatic example is the Haldane model, which demonstrates that a quantum anomalous Hall state can emerge through complex next-nearest-neighbor hopping that breaks time-reversal symmetry\cite{Haldane1988}, or Floquet topological insulators that appear upon laser driving\cite{Oka2019,Rudner2020,McIver2020}. In such systems, topologically protected chiral edge modes form at the boundary and provide unidirectional channels that are resilient to disorder and back-scattering \cite{HasanKane2010,Rechtsman2013,Hong2012,Tiwari2019}. Understanding how these edge states respond to strong optical fields on ultrafast timescales is of foundational importance for condensed-matter physics and prospective optoelectronic or quantum-information applications in Petahertz electronics\cite{Heide2024} and quantum devices\cite{FuKane2008}.

\begin{figure}[ht]
\centering
\includegraphics[width=\linewidth]{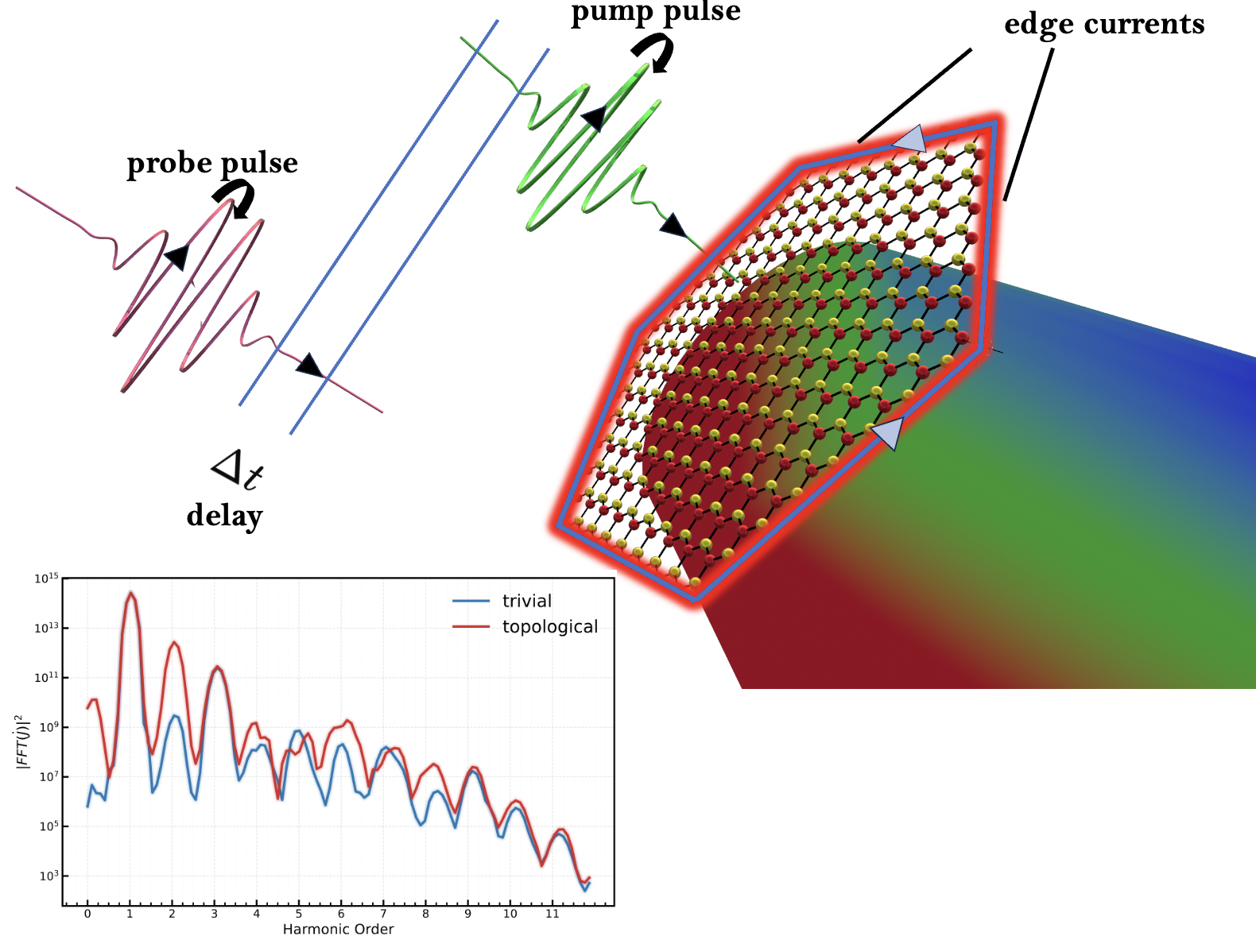}
\caption{Illustration of DCD set-up. A pump pulse excites edge currents in a Haldane nanoflake, and a subsequent delayed ($\Delta t$) probe pulse generates HHG from the current-carrying state, which can be indicative of the pumped topological bulk/edge photocurrents and their ultrafast evolution. Both pump and probe are circular, allowing four configurations of alternating helicities that map the full double circular dichroism (DCD) response. The inset shows typical HHG emission from trivial and topological states in this pump--probe configuration for a specific choice of helicities. The calculations use $\phi=\pi/2$ and $\Delta=0.00628$ a.u. for both phases, with $t_2=0.001$ a.u. for the trivial state and $t_2=0.15$ a.u. for the topological state. By measuring HHG with different helicities and laser parameters, one hopes to probe the coupled bulk and boundary dynamics and extract the topological character of the sample.}
\label{fig:setup}
\end{figure}

High-harmonic generation (HHG) in solids \cite{Ghimire2019,Yue2022b} has recently emerged as a powerful tool for probing electronic structure \cite{PhysRevLett.115.193603,Lanin2017,PhysRevLett.118.087403}, Berry curvature \cite{Luu2018,Lv2021}, and potentially topological invariants in solids on attosecond to femtosecond timescales \cite{Bauer2018,Silva2019,Uzan-Narovlansky2024,Heide2022a,Schmid2021,Baykusheva2021a,Neufeld2023_PRX,Bai2021,Qian2022}. Unlike perturbative nonlinear optics, strong-field HHG provides access to nonperturbative electron trajectories, interband polarization, and light-driven currents. Multiple works have suggested that the harmonic spectrum encodes highly sensitive information about the underlying symmetry and topology of the electronic states \cite{Bauer2018,Vampa2014,Luu2015,PhysRevLett.118.087403,Uzan2020a,Langer2017,Heide2022a,Lv2021,Bharti2024}. In particular, circularly polarized HHG has been predicted to distinguish trivial and topological phases in bulk Haldane systems, revealing contrasts in emission directionality, spectral weight, and helicity selectivity that arise from Berry curvature and anomalous velocity effects \cite{Silva2019,Chacon2020,Jurs2022}. However, most existing HHG studies focus on extended crystals with periodic boundaries, where HHG signatures arise from bulk band geometry and were shown not necessarily linked to topology \cite{Neufeld2023_PRX,Qin2024}. Finite topological systems, such as nanoscale flakes, quantum dots, molecular lattices, or nanoribbons, could introduce new elements to the process via confinement of protected edge states, while the ratio of edge to bulk degrees of freedom can generally be tuned. Here the interplay between surface currents and strong optical driving becomes directly accessible, but the contribution of edges and bulk to HHG, and especially their coupling, is not yet fully understood. Indeed, only few works previously explored the role of topological surface states in HHG \cite{Bauer2018,Jurs2022,Jurs2019,Baykusheva2021a,Baykusheva2021,Schmid2021,Graml2023,Luo2024}, and none in the context of probing laser pumped edge photocurrents. In particular, a better understanding of how to selectively separate HHG signatures emerging from surface/bulk states is needed\cite{Li2026}, as well as from topologically protected surface currents which are ultimately desired for Petahertz optoelectronics\cite{Heide2024} and for exploring topologically-protected physics in novel quantum phases. Schemes for estimating the potential topological aspects of such signatures are also lacking. 

Here we theoretically explore HHG in finite Haldane nanoflakes driven by circularly-polarized pump and probe lasers. We  search for the potential role of topologically-protected edge currents in emitted spectra. We examine harmonic emission amplitudes across various driving strengths, harmonic orders, and the Haldane complex hopping parameter that controls the transition between topologically trivial and nontrivial phases. By fixing the pump helicity and selectively reversing only the probe helicity, we compute harmonic-resolved CD. In select harmonic orders, we find that strong CD arises only in the topological phase, while the trivial phase leads to vanishing CD (in agreement with past works\cite{Chacon2020,Silva2019a}), which could be potentially useful for topological spectroscopy if used with caution (keeping in mind that the result is not general to all harmonics, and likely specific to Chern insulators). Furthermore, by systematically switching the helicity of both the pump and probe, we introduce double circular dichroism (DCD) as a novel probe capable of separating bulk/edge dynamics. We show that in typical conditions DCD contributions from the bulk and edge carry similar magnitudes but opposite signs, and the edge-response arises from the outermost atomic layer of protected modes. DCD can hence be employed for isolating contributions of topological edge currents more selectively than conventional HHG-CD, also since DCD signals vanishes in systems that lack photocurrent-carrying states and are time-reversal symmetric. We also show that DCD has more favorable robustness with respect to system size, gap, and hopping amplitudes, and exhibits favorable reduced sensitivity to bulk/edge interference under certain conditions. By exploring the delay-dependent dynamics we show that DCD signals oscillate on attosecond timescales, which can probe intrinsic pumped photocurrent dynamics as a novel ultrafast spectroscopic tool, e.g. sensitive to photocurrent coherence. Lastly, we find a typical change in behavior of DCD across the topological phase transition: trivial phases exhibit stable non-oscillatory DCD, while topological phases are very sensitive and show strong oscillatory patterns. This feature might be employed for probing topological features in system with broken time-reversal symmetry by tuning the material system via strain engineering, doping, or laser dressing. Our work establishes DCD as an all-optical approach for probing quantum systems with broken symmetry (e.g. topological and/or chiral), and should open new routes for extended DCD optical spectroscopies in other avenues such as photoemission and absorption.

\section{Results} 

\subsection{Haldane nano-flakes} 
\begin{figure}[t]
    \centering
    \includegraphics[width=0.5\textwidth]{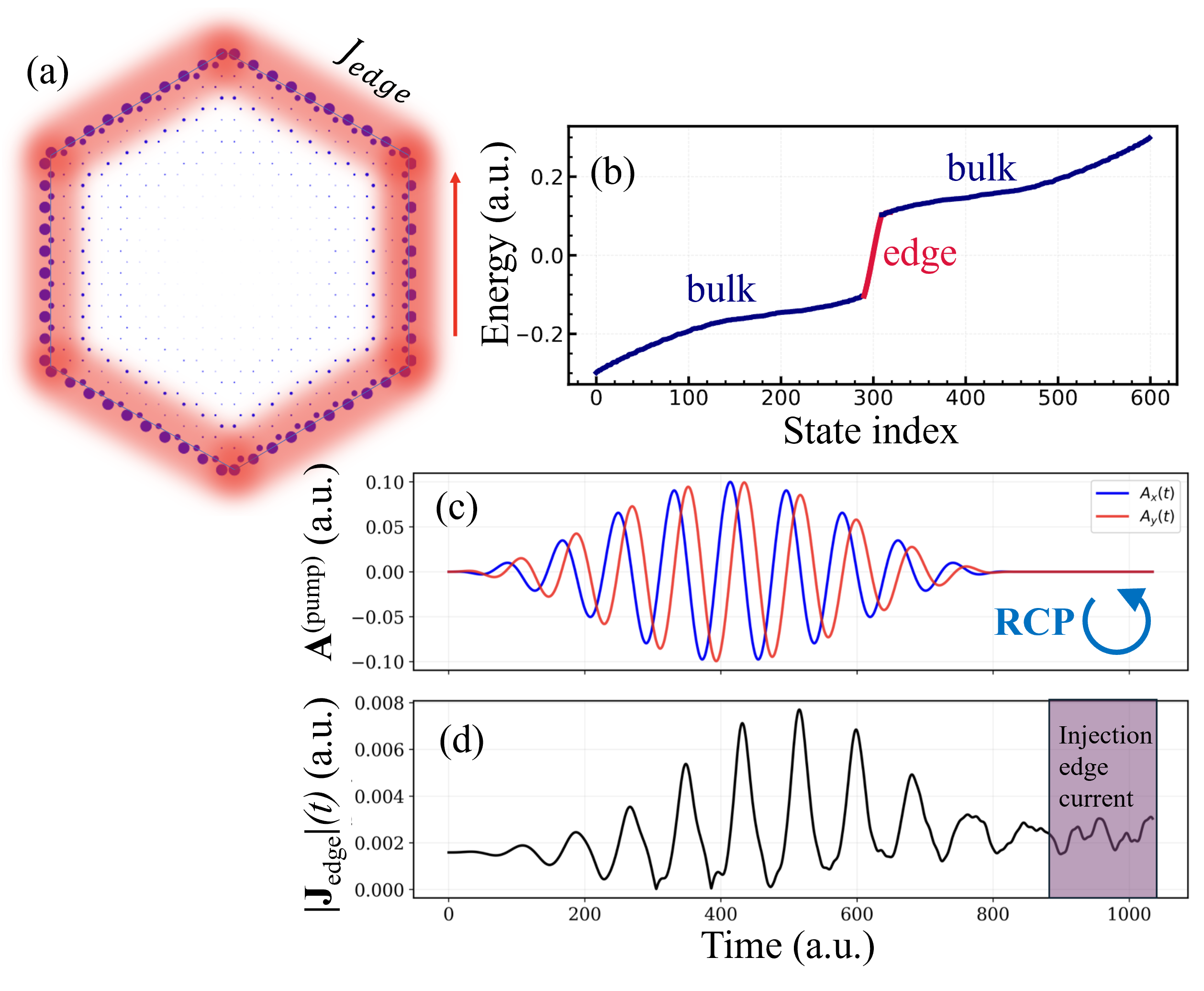}

    \caption{(a) Hexagonal 2D flake of size $m=5$ showing atomic positions arranged on a honeycomb lattice and edge currents animated in red. The site size is proportional to its occupation in this illustrated helical TPSS. (b) Energy spectrum eigenmodes for this system vs state index for parameters $\phi = \frac{\pi}{2}$, $M = 0.01$, $t_2 = 0.03$, and $t_1 = -0.1$. (c) Pump pulse profile with the zero-field region and (d) the resulting edge current with the temporal region where the average edge current is calculated highlighted.}
    \label{fig:flake_combined}
\end{figure}

Let us begin by introducing our theoretical approach and methodology. We consider a finite hexagonal nanoflake constructed from the Haldane model on a honeycomb lattice (see illustration in Fig.~\ref{fig:setup}). The system consists of A and B sublattice sites with finite boundary conditions, forming a well-defined perimeter where chiral edge modes can appear. A representative flake geometry is shown in Fig.~\ref{fig:flake_combined}(a). Unlike in the periodic system, translational symmetry is absent and the electronic structure is described by a discrete set of eigenstates whose spatial localization---edge versus bulk---plays a central role in the nonlinear optical response. The finite flake is governed by the lattice Hamiltonian,
\begin{equation}
H = -t_1 \sum_{\langle i,j \rangle} c_i^\dagger c_j 
+ t_2 \sum_{\langle\!\langle i,j \rangle\!\rangle} 
e^{i \nu_{ij} \phi} c_i^\dagger c_j
+ M \sum_i \xi_i\, c_i^\dagger c_i ,
\end{equation}
where $t_1$ is the nearest-neighbor (NN) hopping amplitude, $t_2 e^{i\phi}$ is the complex next-nearest-neighbor hopping (NNN) term that breaks time-reversal symmetry, $\nu_{ij}=\pm 1$ denotes the handedness of the NNN loop, and $M$ is a staggered on-site potential with $\xi_i=\pm1$ on the A/B sublattices with $i$ and $j$ site indices in the flake. For suitable choices of $(t_2,\phi,M)$, the system lies in a Chern-insulating regime, supporting unidirectional protected boundary states confined to the flake edges (so-called topologically-protected surface states (TPSS)). The flake geometry is chosen to be a six-fold rotationally-symmetric hexagon, respecting the individual unit-cell symmetry, while in this work we fix the flake size to $m=5$ that shows sufficient edge-state localization in the topological phase (see Fig.\ref{fig:flake_combined} and SI for size-dependency tests).

Diagonalization of this Hamiltonian provides its eigenvalues $E_n$ and eigenstates $\psi_n(i)$. Figure ~\ref{fig:flake_combined}(b) presents the three classes of states that appear: (i) valence-band bulk states, (ii) conduction-band bulk states, and (iii) a connected set of in-gap edge states. These edge states form a nearly linear sequence (with linear dispersion) bridging the valence and conduction manifolds and exhibit strong spatial localization at the outer boundary of the hexagon (see illustration in Fig.~\ref{fig:flake_combined}(b)). The chiral nature of the edge states reflects circulating currents with a directionality set by the complex phase of $t_2$ --- a direct manifestation of the broken inversion and time-reversal symmetries of the Haldane Hamiltonian and the sign of the Chern index. Bulk-edge state coupling is included in the Hamiltonian by construction.

\subsection{Laser pumped helical photocurrents} 
We next explore the ultrafast electronic response of the Haldane nano-flakes to intense pump laser driving. We perform real-time propagation of all occupied single-particle eigenstates (with a 4th order Runge-Kutta scheme) under a circularly polarized pump field. The dynamics follow the time-dependent Schrödinger equation:
\begin{equation}
i\,\partial_t |\psi(t)\rangle = H(t)\,|\psi(t)\rangle ,
\end{equation}
where $H(t)$ is obtained from the static Hamiltonian via Peierls substitution. For a site pair $(i,j)$ separated by displacement $(\Delta x_{ij}, \Delta y_{ij})$, the field modifies the hopping amplitude as
\begin{equation}
H_{ij}(t) = H_{ij}\,\exp\!\left[-i\left(\Delta x_{ij} A_x(t) + \Delta y_{ij} A_y(t)\right)\right],
\end{equation}
with $(A_x(t),A_y(t))$ the components of the pump vector potential.  
We employ the dipole approximation and use a pulse with a $\sin^2$ envelope of duration $2\pi n_0 / \omega_0$, given by
\begin{equation}
\mathbf{A}^{(pump)}(t) = A_0^{(pump)} \, \sin^2\!\Big(\frac{\omega_0 t}{2 n_0}\Big) \, \sin(\omega_0 t) \, \mathbf{\hat{e}}
\end{equation}
Here $n_0 = 10$, the field amplitude $A_0^{(pump)}$ is a tunable parameter that is connected to the laser peak power, and the angular frequency is $\omega_0 = 0.0759$ a.u. (corresponding to a wavelength $\lambda_0 \approx 600\,\mathrm{nm}$), which we have found to yield clean harmonic spectra (see example HHG spectrum in Fig. \ref{fig:setup}). $\mathbf{\hat{e}}$ is a circularly-polarized unit vector with varying helicity, ($\mathbf{\hat{e}_{R/L}} = (\hat{x} \pm i\hat{y})/\sqrt(2)$).
The pulse is followed by a field-free interval of $\sim$10 fs where we extract persistent circulating currents generated by the pump (see Fig.\ref{fig:flake_combined}(c)). Note that no dephasing or decoherence channels are included in our simulations (apart from bulk-edge coupling), such that induced excitations are expected to remain coherent. This should be a good approximation in topological insulators with long coherence times \cite{Soifer2019,Schmid2021,Lively2024,Mitra2024}, certainly longer than few tens of femtoseconds - the timescale of our simulations (though in HHG this is today a hotly debated topic with ongoing research\cite{Cardenas2026EffectsSolids,Herling2026}). 

The laser-driven current is computed as:
\begin{equation}
\textbf{J}_\alpha(t) = 
\sum_{n}\sum_{i,j}
\left(-i\, \psi_n^*(i,t)\,H_{ij}(t)\,\psi_n(j,t)\right)
\Delta \textbf{r}_{ij}
\end{equation}
where $\Delta \textbf{r}_{ij}\in\{\Delta x_{ij},\Delta y_{ij}\}$ selects the Cartesian displacement component. In this approach, the time-dependent electronic currents can be evaluated both over the entire flake, or in specific regions. Using the identified boundary sites of TPSS (see Fig.\ref{fig:flake_combined}(a)) we compute boundary-resolved currents that quantify circulating motion within the selected boundary region via:
\begin{equation}
\textbf{J}^{\mathrm{edge}}_\alpha(t)
= \sum_{i,j \in \mathrm{edge}} 
\left(-i\, \psi^*(i,t)\,H_{ij}(t)\,\psi(j,t)\right)
\Delta \textbf{r}_{ij},
\end{equation}
\noindent Figure \ref{fig:flake_combined}(c) shows a representative example of the pump-induced response driven by a left circularly-polarized (LCP) pulse.

\begin{figure*}[t]
\centering

    \includegraphics[width=0.8\textwidth]{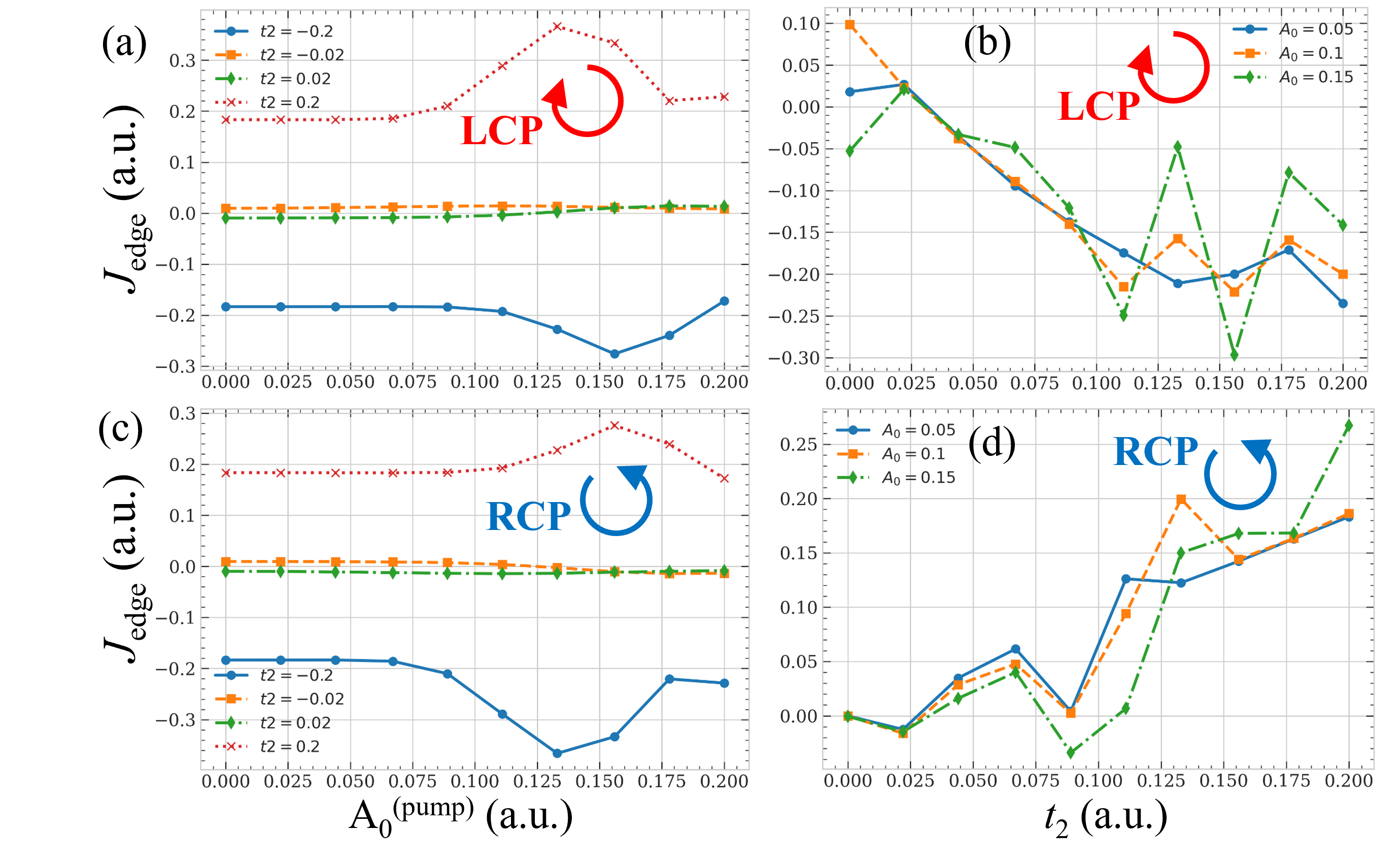}

\caption{Laser pump tunable edge photocurrents, $J_{edge}$, in different laser-matter conditions. (a,c) Photocurrent generation induced by a circular pump with left circular polarization (LCP) in (a), and right CP (RCP) in (c), calculated vs. different pump field amplitudes, and for various values of the NNN hopping $t_2$. The topological phase arises for $t_2=\pm0.2$ a.u., while a trivial phase emerges for $t_2=\pm0.02$ a.u. (b,d) Similar to (a,c) but vs the $t_2$ parameter and for a few different values of $A_0^{(pump)}$.}

\label{fig:currents_combined}
\end{figure*}

By averaging $\textbf{J}(t)$ in the field-free region (see illustration in Fig. \ref{fig:flake_combined}(c)), we extract photocurrents generated by the pump (with a methodology similar to that in refs. \cite{Neufeld2021a,Galler2025}), which could serve as an ultrafast probe of the topological character of the system \cite{Weitz2024,Lesko2025}. We perform this analysis across a wide range of parameters, including complex NNN hopping, the pump amplitude, and pump helicity. The resulting induced photocurrents in Fig. \ref{fig:currents_combined} reveal a clear enhancement in the topological regime --- light-driven edge currents are more easily manipulated and excited through the TPSS compared to the trivial system (Fig.\ref{fig:currents_combined})). Importantly, we observe a dichroic effect where flipping the pump helicity changes the induced photocurrent amplitude (not only sign), which is a result of the broken time-reversal and inversion symmetries in the Haldane Hamiltonian. Flipping both pump helicity and sign of NNN term reverses the response, as expected (e.g. compare red/blue lines in Fig.\ref{fig:currents_combined}(a) and (c)).

\subsection{Pump--Probe HHG CD and DCD} 
Next, we explore potential signatures of light-manipulated photocurrents in HHG. The circular pump drives photocurrents, and a delayed intense circular probe pulse generates HHG from the current-carrying states (analogously to ring current spectroscopies \cite{Eckart2018,Neufeld2019b,Moitra2025}). This is simulated by adding to the total laser field an additional probe pulse having the same carrier frequency $\omega_0$ and duration, given by  
\begin{equation}
\resizebox{0.95\linewidth}{!}{$
\textbf{A}^{(\mathrm{probe})}(t)
=
A_0^{(\mathrm{probe})}
\sin^2\!\left(
 \frac{\omega_0 (t-\tau)}{2 n_0}
\right)
\cos(\omega_0 (t-\tau))\mathbf{\hat{e}}
$}
\end{equation}

\noindent where $\tau$ is the pump-probe delay, which is taken to be longer than the pump pulse duration ($\tau\sim$26 fs). This value was selected to prevent temporal overlap of the two approximately 20-fs pulses and is not a unique or optimized. These pulse durations were chosen as they reflect experimentally accessible conditions (note that increasing the pump duration is equivalent to increasing its peak power in terms of the photocurrent pumping it induces). The inter-pulse delay dependence will be explored below, while the majority of the text fixes this value to  $\sim$6 fs unless stated otherwise. HHG emission is calculated by Fourier transforming the time-derivative of the total current, while individual harmonic yields are extracted by integration around each order. HHG generated solely by the probe is calculated by filtering the current temporally around in the probe's temporal region, mimicking experimental conditions where the pump and probe's individual responses can be separated\cite{Baykusheva2019,Mitra2024,Tyulnev2024}. Circular dichroism (CD) for each harmonic is defined using the HHG intensities $I_n^{(-,+)}$ and $I_n^{(-,-)}$ for right- and left-circular probe fields (with fixed pump helicity):

\begin{equation}
\mathrm{CD}_n
=
\frac{
 I_n^{(-,+)}
 -
 I_n^{(-,-)}
}{
 I_n^{(-,+)}
 +
 I_n^{(-,-)}
}.
\end{equation}
where $I_n^{(+/-)}$ is the integrated $n$'th harmonic yield for a given pump-probe helicity configuration. The first component sign in the $(\pm,\pm)$ notation corresponds to the helicity of the pump, while the second corresponds to the helicity of the probe. Reversing either the probe or pump helicities strongly modulates the HHG emission, which is fundamentally a result of broken time-reversal and inversion symmetries in the Haldane phase.

\begin{figure*}[t]
    \centering
    \includegraphics[width=0.9\textwidth]{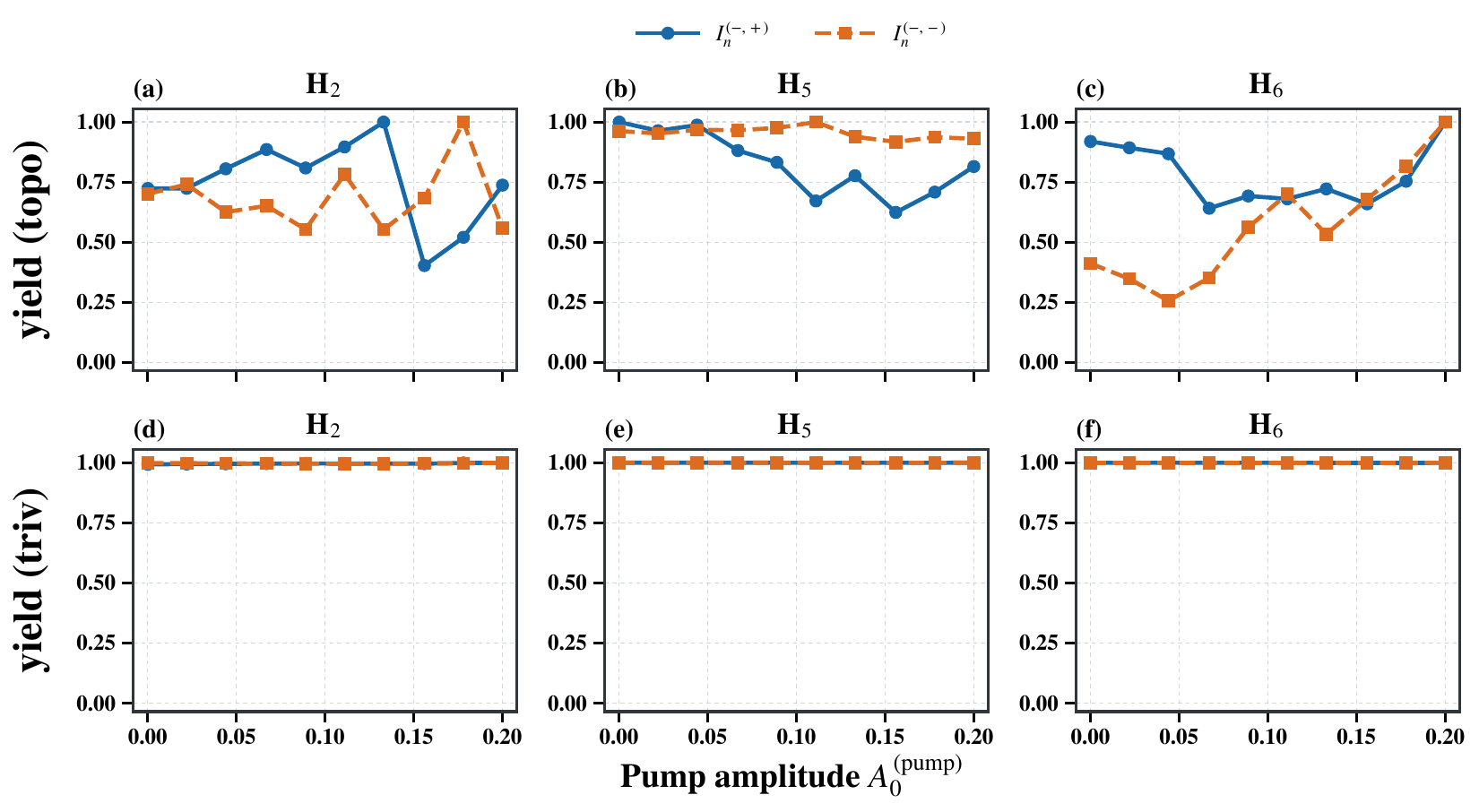}
\caption{
        Normalized integrated harmonic yields driven by circularly polarized pump-probe pulses.
        (a-c) HHG yields driven by right/left circular probe pulses with a fixed LCP helicity for the pump pulse, for $t_2 = 0.15$ a.u. and $\Delta = 0.1$ for select harmonic orders. (d-f) Same as (a-c) but for the trivial phase case ($t_2 = 0.001$ a.u., $\Delta = 0.00628 $ a.u.). The harmonics orders 2,5,6 were chosen since in these harmonics the HHG CD for the trivial phase nearly vanishes. All remaining harmonic data can be found in the Supplementary Information (SI).}
    \label{fig:hhg_yields_topo_trivial_1}
\end{figure*}

Figure~\ref{fig:hhg_yields_topo_trivial_1} presents probe–induced HHG yields for select harmonics as a function of the pump power, while the probe power is fixed, for two representative values of the hopping parameter, $t_2=-0.001$ a.u. (trivial), and $t_2=-0.15$ a.u. (topological). For the topological phase, we find that harmonic yields strongly depend on $A_0^{(pump)}$, suggesting that edge photocurrents and their coupling to the interior influence HHG. This is perhaps not surprising given analogous results were seen in pumped solid and molecular systems\cite{Hohenleutner2015,Heide2022,Hamer2022,Wanie2024}. Strongly oscillatory features likely connect to an interplay of probe pulse induced dynamics and hot carrier effects from the pump, where Pauli blocking can also reduce HHG yields of pre-pumped systems \cite{Heide2022,DeKeijzer2024,Chen2025a}. What is crucial is that in this case photocurrents populate the TPSS, as discovered by the pump-only analysis above. The HHG response should therefore provide sensitivity to TPSS dynamics.

An interesting result arises if considering the trivial case. Here much milder responses are observed, where HHG yields are roughly independent of the pump power. This is somewhat expected due to the absence of TPSS in the trivial phase leading to weaker current-carrying edge states (as in Fig. \ref{fig:currents_combined}). Still, we recall that photocurrents induced by the pump can also circulate in the bulk of the trivial phase, and in some harmonics bulk response can dominate. Indeed, analysis of all harmonic orders (see SI) shows that some harmonics exhibit vanishing CD, and some do not. Nevertheless, by pre-selecting harmonics that are particularly sensitive to the presence of the boundary, and by tuning the pump power, one could in principle identify the presence of TPSS and topological attributes in the system. Note that this is a more robust implementation compared to earlier proposals in the sense that the probe pulse that generates HHG maintains a fixed power; hence, changes in HHG yield by construction arise from changes in pumped photocurrents. We also note that we tested several sizes of flakes up to $m=9$, and largely similar results were observed throughout (see SI). We expect this scheme to be applicable in Chern insulators or other systems with broken symmetry.

\begin{figure*}[t]
    \centering
    \includegraphics[width=1\textwidth]{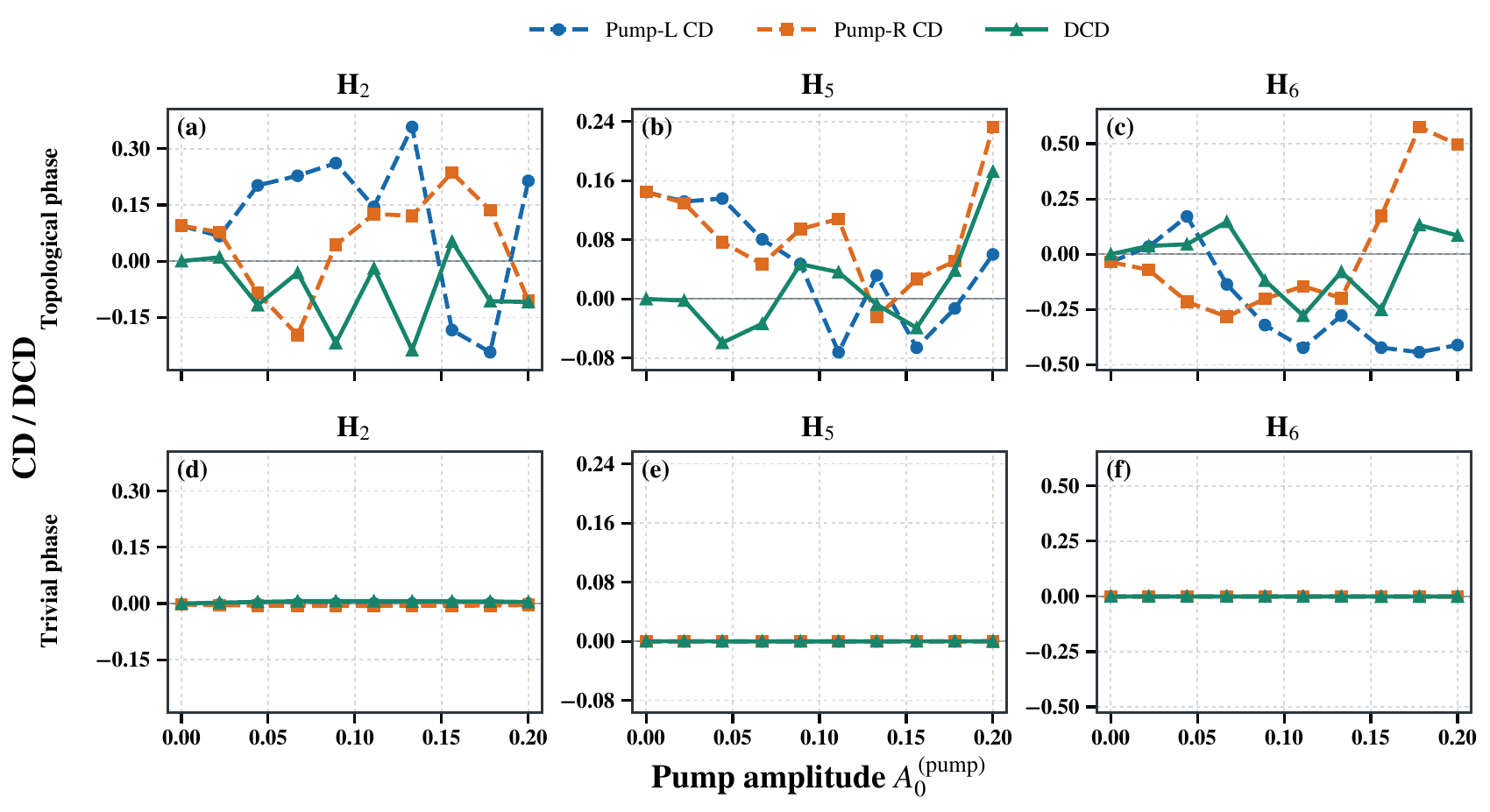}
    
    \newcommand{\w}{0.27\textwidth}      
    \newcommand{\colsep}{0.015\textwidth} %
    \caption{
        Corresponding HHG CD and DCD, defined as $|CD_R|-|CD_L|$, in similar conditions to Fig. \ref{fig:hhg_yields_topo_trivial_1}.}
    \label{fig:cd_A0_topo_trivial_1}
\end{figure*}

Figure~\ref{fig:cd_A0_topo_trivial_1} presents normalized HHG CD of select harmonics as a function of the pump amplitude for the trivial and topological cases. For the trivial case, the CD vanishes for these harmonics, indicating absence of chiral asymmetry (expected from Fig. \ref{fig:hhg_yields_topo_trivial_1} showing similar HHG amplitudes for RCP/LCP probe helicities). On the other hand, the topological phase exhibits strong CD in agreement with previous predictions \cite{Chacon2020,Silva2019a}. As $A_0^{pump}$ increases, the CD magnitude generally grows, though we note that this behavior is non-monotonic and is also prone to sign changes for sufficiently intense pumps, indicating an intricate highly nonlinear relationship between pump-induced photocurrents and HHG. Indeed, this oscillatory behavior is not apparent in induced edge-currents by the pump alone in Fig. \ref{fig:currents_combined}. The onset and strength of the CD also varies slightly across different harmonics (see full data for all harmonics in SI), reflecting additional sensitivity to the chiral dynamics. Note that the sign changes with respect to pump power indicate that CD on its own is not necessarily a strong indicator of the topological phase, as proposed in ref. \cite{Neufeld2023_PRX}, since it implies that under certain laser driving regimes the CD is not directly linked to the Chern number, or that link it too nonlinear and convoluted. However, it is certainly strongly impacted by both the topological phase and induced photocurrents, which might be extractable by more sophisticated observables. We also caution again that we display in Figure~\ref{fig:cd_A0_topo_trivial_1} only select harmonics for which the CD vanishes in the trivial case, while in the SI some opposite cases can be found.

We next compute the differential circular dichroism, i.e. the normalized difference between HHG CD (obtained upon reversing the probe helicity) when the pump is left- versus right-circularly polarized. We denote this double-CD observable as DCD:
\begin{equation}
\mathrm{DCD}_n
=
\left|CD_n^{(\mathrm{pump\,R})}\right|-\left|CD_n^{(\mathrm{pump\,L})}\right|,
\end{equation}
\noindent
where $CD_n^{(\mathrm{pump\,R})}$ and $CD_n^{(\mathrm{pump\,L})}$ are the CD values of the $n$-th harmonic when the pump is right- and left-circularly polarized, respectively. Since each CD is normalized in $[-1,1]$, the DCD is similarly bounded. Overall, four separate harmonic yields enter the DCD construction ($(+,+)$, $(-,-)$, $(+,-)$, and $(-,+)$), and this definition upholds two important limits: (i) the DCD inherently vanishes in systems respecting time-reversal symmetry (see SI); (ii) the DCD vanishes as the pump power ($A_0^{(pump)}=0$) is reduced to zero (see Fig. \ref{fig:currents_combined}), meaning it is particularly sensitive to pumped photocurrents. We are motivated to examine the DCD since preceding results indicate that pump-induced chiral dynamics are imprinted on the emitted harmonics, while the pumped photocurrents themselves were dichroic in Fig. \ref{fig:currents_combined}. Note that there also exist two alternative ways of defining DCD that are unique (based on the combinatorics of the four individual HHG yields with various helicities), but here we use the above definition due to its intuitiveness and upholding the proper limits (which other definitions do not necessarily keep). A nonzero DCD therefore evaluates a pump-helicity-dependent imbalance in CD magnitude, which is unique to broken symmetry systems, as in systems respecting time-reversal, inversion, or other symmetries\cite{Neufeld2019}. This naturally motivates exploring DCD, since it might be effective for topological systems like Chern insulators and topological magnets\cite{Bernevig2022}, unlike regular CD that also appears in trivial systems if they have broken symmetries, or might not appear in some topological systems\cite{Heide2022a}.

Figure \ref{fig:cd_A0_topo_trivial_1} presents resulting DCD vs the pump pulse power. Note that the select harmonic orders chosen in Fig. \ref{fig:cd_A0_topo_trivial_1} show vanishing DCD in the trivial case, but this is not a generic feature since for small but nonzero $t_2$ the system can be in a trivial phase and still exhibit broken time-reversal symmetry. Indeed, in the SI we show that some harmonics in the trivial case still exhibit DCD due to pumped photocurrents in their bulk. Generally, the DCD response in the topological phase is highly nonlinear and nonperturbative, which arises in this regime since both pump and probe beams are very intense. In particular, an intense pump is required in these conditions to effectively pump a photocurrent into the edge. Thus, DCD is appealing for temporally resolving ultrafast photocurrent dynamics in a pump-probe scheme. In the SI we further show that the DCD responses are more stable across the flake sizes compared to typical CD, making it robust for ultrafast spectroscopy. 

\begin{figure*}[t]
    \centering
    \includegraphics[width=\textwidth]{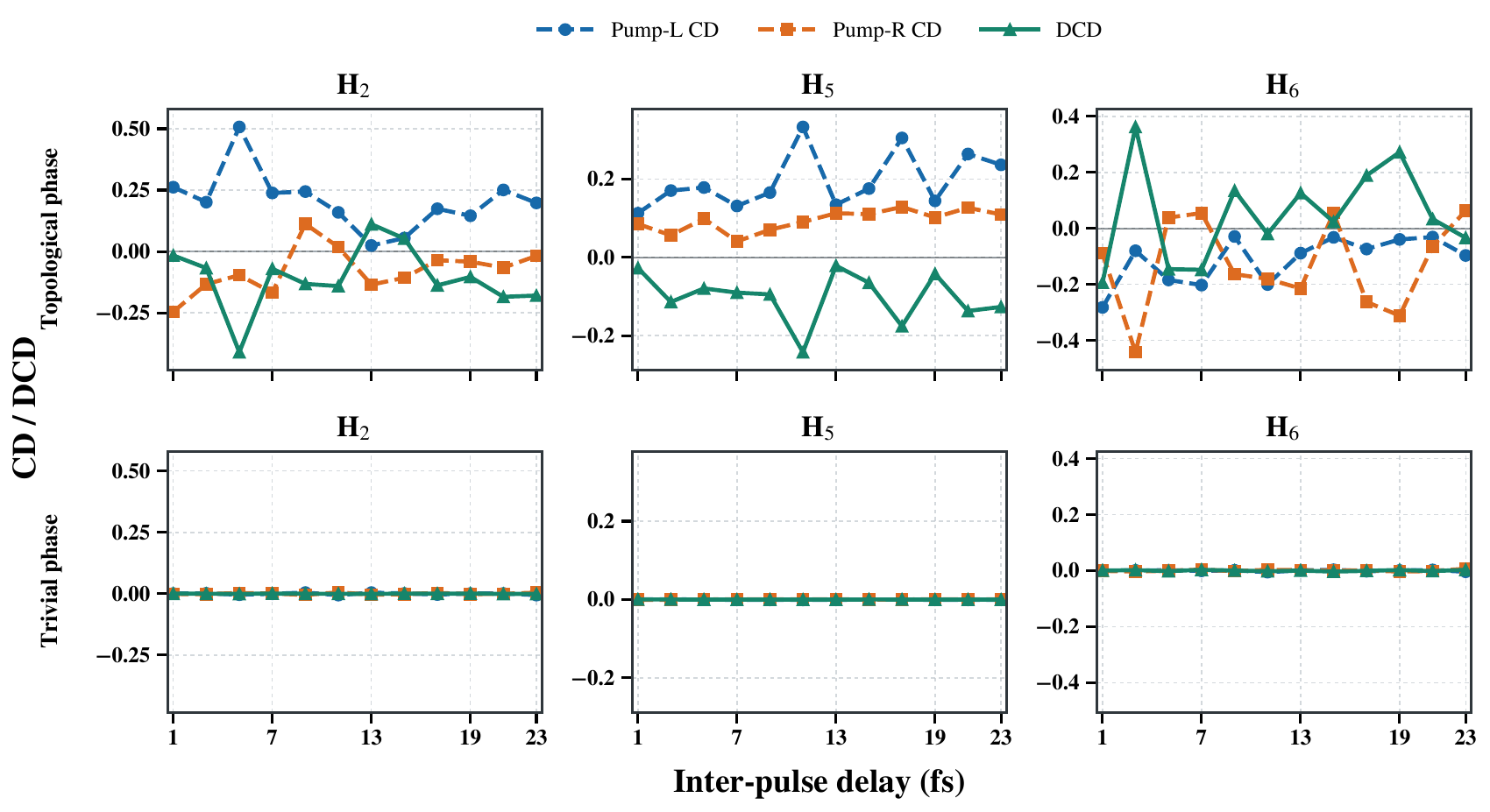}
    \caption{Delay dependence of the pump-R CD, pump-L CD, and DCD for harmonic orders 2, 5, and 6 in the topological (top row) and trivial (bottom row) phases. The horizontal axis gives the field-free interval between the pump and probe pulses. Calculations use $A_0^{(\mathrm{pump})}=0.1$ a.u., $A_0^{(\mathrm{probe})}=0.15$ a.u., $m=4$, and $\phi=\pi/2$, with $(t_2,\Delta)=(0.15,0.1)$ a.u. for the topological phase and $(0.001,0.00628297)$ a.u. for the trivial phase. The remaining studied harmonic orders are shown in the SI.}
    \label{fig:delay_cd_dcd}
\end{figure*}

To understand its potential ultrafast capabilities as a spectroscopic probe, we explore the delay-dependent signal. Figure~\ref{fig:delay_cd_dcd} follows the helicity-resolved response vs pump--probe delay. The dynamics are shown to be highly oscillatory on attosecond timescales, indicating that the DCD carries information about the evolution of the pumped photocurrent state and its coherence and could be used to probe topological photocurrent dynamics. In unique cases, DCD can even incur sign changes over the temporal evolution as in H6 in  Fig. ~\ref{fig:delay_cd_dcd}, indicating the nonlinearity of the probe and the fact the relationship to the pump pulse is not one-to-one. In the pre-selected harmonics 2,5,6, the trivial phase shows no response, which might be useful for probing Chern-like physics and its topological attributes, though we note that in the SI we show that for other harmonics the trivial system can still show some delay dependence as well as non-vanishing DCD. This is a result of mostly bulk pumped photocurrents, while the delay dependence is on average much weaker in the trivial phase, attributed to its lack of conducting edge modes.

\begin{figure*}[t]
    \centering
    \includegraphics[width=0.9\textwidth]{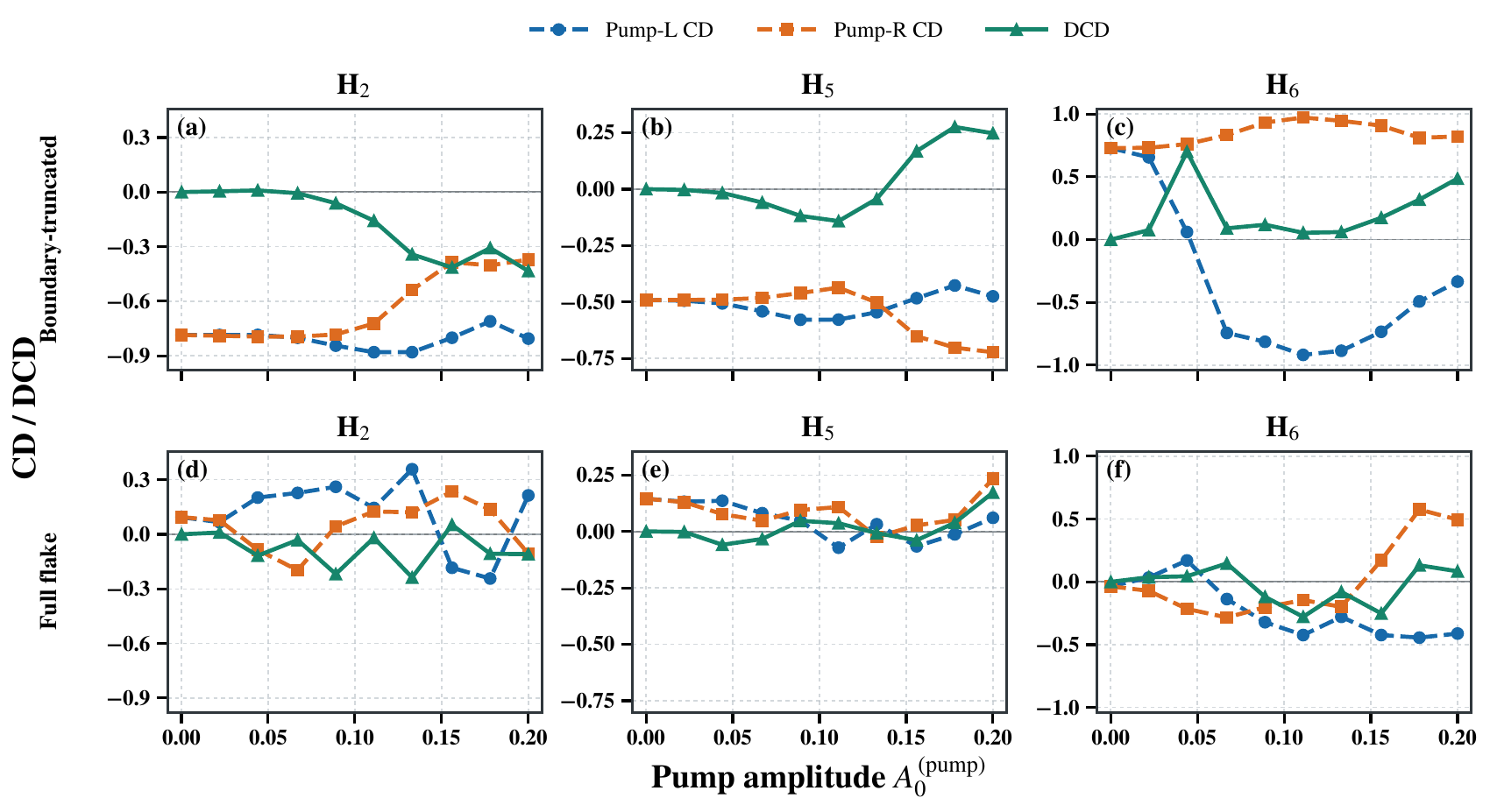}

    \caption{
        Corresponding HHG CD and DCD in similar conditions to Fig. \ref{fig:cd_A0_topo_trivial_1} for the topological phase, but where (a-c) is calculated after removing the outer atomic layers, and (d-f) is the coherent full-flake response. Their comparison quantifies contributions of edge state physics to DCD (also see SI).
    }
    \label{fig:no_edge_comparison}
\end{figure*}

\subsection{Bulk-edge separation in DCD} 
One key challenge in HHG-based probes of topological systems is identifying the microscopic origin of observed signals. This is a challenge even in photocurrent spectroscopies\cite{Soifer2019,Weitz2024}. In particular, it is often unclear whether a given response arises from bulk dynamics, circulating edge currents, intrinsic topological properties, or from pump-induced population imbalances \cite{sato2019light}. Such a separation is crucial in order to explore topologically-protected state dynamics in intrinsic-topological and Floquet-topological systems. 

To assess the sensitivity of HHG to edge-associated dynamics, we compare the coherent response of the full flake with that of a truncated flake in which the current associated with the outer boundary layer of atoms are removed (which contains the dominant part of the exponentially-localized TPSS). All full-flake spectra are calculated from the total time-dependent current and therefore include bulk--edge coupling and interferences. In the SI we demonstrate that further removing the 2nd and 3rd layers keeps the bulk-DCD signals mostly stable, indicating that indeed the last atomic layer carries the key contribution to the surface-localized response and that DCD is a surface specific probe.

We find that HHG CD is typically enhanced (in absolute values) when the outer boundary layers are removed (see Fig. \ref{fig:cd_A0_topo_trivial_1}) compared to the full flake. For several harmonics the boundary-truncated configuration exhibits CD amplitudes that exceed those of the full system by factors of two or more (see also full data set in SI). This confirms that bulk electron dynamics can generate a strong dichroic response (as previously reported\cite{Silva2019a,Chacon2020}), while pumped edge states partially suppresses the net CD through oppositely-signed competing contributions. Thus, the coherent CD is highly sensitive to boundary-associated degrees of freedom and their coupling to the interior. A similar interplay has been identified in confined topological systems, where edge contributions can cancel or compensate each other\cite{Bauer2018}. In particular, recent work has shown that in isolated quantum Hall systems bulk CD is counterbalanced by edge contributions, and that resolving specific sectors is essential for interpreting dichroic signals correctly \cite{unal2025circular}. Our simulations extend this notion to highly nonlinear and non-perturbative responses.

\begin{figure*}[t]
\centering
     \captionsetup[subfigure]{labelfont={bf,Large}}
    \includegraphics[width=0.99\textwidth]{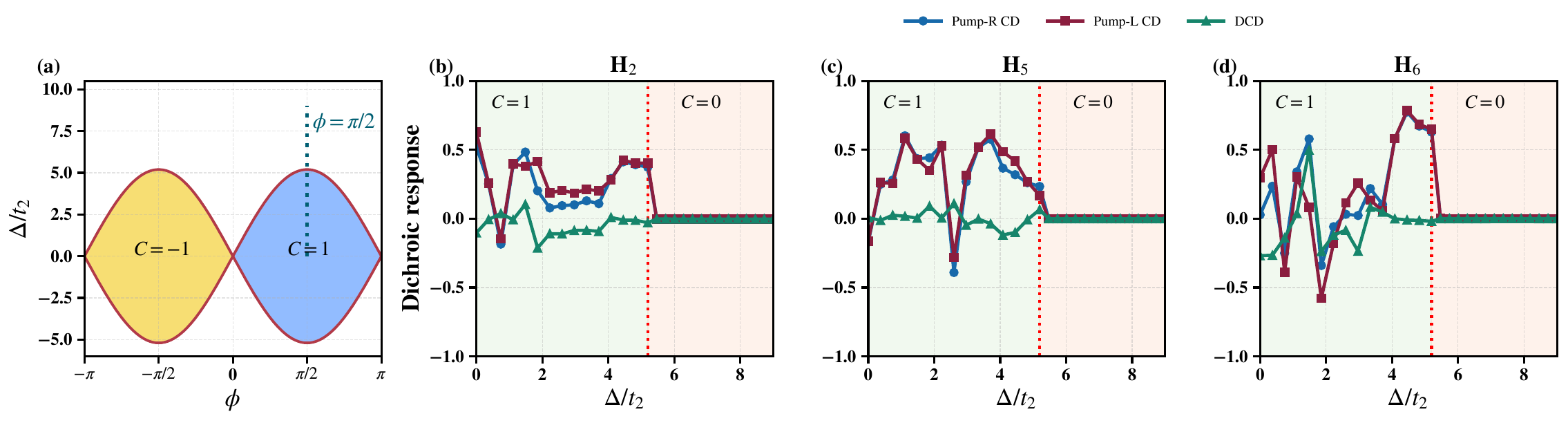}
%
%

    \caption{
        Dependence of HHG CD and DCD on Haldane phase. (a) Phase diagram of Haldane model. The scanned line across the phase transition is indicated in dashed blue line (for $\phi=\pi/2$ and $A_0=0.089$). (b-d) HHG CD and DCD for select harmonic orders across the phase transition labeled in (a). All other harmonic orders are delegated to the SI.
    }
    \label{fig:phase_diagram_dcd}
\end{figure*}

Calculated DCD values still reflect this trend, but are normalized to zero at the non-pumped flake geometries, which therefore makes them selective to light-manipulated photocurrent dynamics. That is, since CD is generally nonzero across topological bulk states even without pre-pumping, it is hard to employ it to track light-induced phases\cite{McIver2020,Rudner2020,Neufeld2025b,Mitra2024,Oka2019} and dynamics\cite{Tyulnev2024,Lively2024,Chen2025a,Heide2022}. Moreover, CD could be nonzero also in trivial phases with broken symmetries. DCD, however, is a direct result of helicity-selective pumped photocurrents, which is generically a signature of broken time reversal symmetry that is ubiquitous in Chern insulators, topological magnets\cite{Bernevig2022}, and Floquet topological systems\cite{McIver2020,Rudner2020,Oka2019,Weitz2024,Lesko2025}. In the SI we explicitly analyze the role of the bulk/edge interference terms and show that they are highly influential in typical HHG, but becomes less crucial in DCD due to cancellations, which endow DCD added robustness.
In this light, our results suggest that DCD can serve as a potential optical probe (either in HHG or other schemes) for separating bulk/edge responses of induced photocurrents, since it is both sensitive to the light-driven current-carrying states, as well as usually arises in similar magnitude and opposite signs from the bulk and edge. Hence, measuring DCD with surface-specific probes (for instance in reflection geometries, or by other means of separating responses of different crystal regions) should allow isolating contributions to photocurrents from TPSS. The sharply different amplitude scaling of the bulk/edge states to DCD (seen in Fig. \ref{fig:cd_A0_topo_trivial_1}) is also a prominent signature that could provide separation via multi-dimensional spectroscopy.  

\subsection{DCD across topological phase transitions} 
Lastly, we explore HHG DCD across the topological phase transition. Up to now we compared few select cases in trivial and topological phases. More generally, by tuning the ratio between the onsite energies (the gap $\Delta$) and the amplitude and phase of $t_2$ one obtains a broad phase diagram for the Haldane model (see Fig.~\ref{fig:phase_diagram_dcd}(a)). We fix the complex hopping phase to $\pi/2$ and vary the ratio of $\Delta/|t_2|$ such that the system transits from the topological to trivial phase continuously (along the dashed blue line in Fig.~\ref{fig:phase_diagram_dcd}(a)). At each point along the diagram we calculate pump-probe HHG CD and DCD, while fixing the pump and probe powers. Data for select harmonics is shown in Fig.~\ref{fig:phase_diagram_dcd}(b--d), while all other harmonic orders are delegated to the SI. From this simulation we observe a strong oscillatory response of the CD and DCD in the topological region, meaning that those quantities are highly sensitive to the system parameters. Notably, this response is much more oscillatory compared to that shown in the past in similar systems in the bulk\cite{Chacon2020}, which may reflect the coherent interplay of boundary-associated and interior dynamics in the finite flake. The DCD is likely oscillatory, but a little more stable compared to regular CD that can have very sharp peaks and sign changes for varying $\Delta/|t_2|$. This indicates DCD might be more well-suited for probing topological properties in broken symmetry system.

On the other hand, one clear result arising is the generally small CD in the topologically trivial region. This region does not show oscillations in the HHG CD/DCD at all. This is inline with results presented above, and suggests that a potential signature of a topological phase transition in HHG: a sharp shift from oscillatory to stable response with respect to the system parameters. The parameters might be experientially tuned via doping\cite{Heide2022a}, or possibly via laser driving\cite{Tyulnev2024,Mitra2024,Tyulnev2024,Neufeld2023} or strain to manipulate the hopping amplitudes\cite{Su2026,Nebgen2026}. We should note that this result again differs from prior results in the bulk of Haldane flakes\cite{Chacon2020}, where contrarily much more oscillatory behavior was seen in the trivial region. Together with the results above, this indicates that HHG CD in Chern insulators (and likely also other topological phases) is highly sensitive to associated TPSS and their coherent coupling to the bulk. In the SI we show that all harmonics emitted from this system generally display this type of behavior, meaning that at least in the case of Chern insulators we expect DCD to provide a strong signature of surface topological photocurrents.

\section{Summary and outlook} 
We theoretically and numerically explored HHG from 2D Haldane nano-flakes in a pump probe set-up and varying laser-matter parameter regimes. We proposed double circular dichroism (DCD) as a new observable in nonlinear strong-field spectroscopy of topological matter, where one evaluates the differential CD by inverting both probe and pump helicities. Unlike conventional CD that probes the helicity sensitivity of a single driving field, DCD exploits the combined helicity dependence of pump and probe, which is shown sensitive to light-induced photocurrents. We showed that DCD enables disentangling optically-induced photocurrent responses due to the bulk and edge leading to similar magnitude, but oppositely signed, DCD, while the DCD vanishes in non-pumped conditions and in quantum phases without broken symmetry. We also demonstrated that bulk and edge contributions to DCD scale differently with the pump amplitude, which serves as another separation mechanism. We clarified that edge-contributed DCD is directly associated with the last atomic layer of edge-localized TPSS, and DCD in general has improved robustness to bulk/edge interference effects. With pump-probe delay-dependent simulations, we found DCD to oscillate on attosecond timescales nonlinearly, likely corresponding to coherence of pumped photocurrents in the edge and bulk regions, showing its potential for attosecond topological spectroscopy.

Simulations across the topological phase transition and different flake sizes showed that DCD exhibits improved stability and robustness compared to typical CD, and might be an indicator of topological phase transitions via a change in oscillatory patterns across the transition (which vanish in the topologically trivial region). Such predictions might be tested by doping\cite{Heide2022a}, or possibly via laser driving\cite{Tyulnev2024,Mitra2024,Neufeld2023} or strain to manipulate the hopping amplitudes\cite{Su2026,Nebgen2026}. It remains currently unclear if similar features will also arise in regular topological insulators, while we expect them to appear in systems with broken time-reversal symmetry such as magnetic topological insulators\cite{Bernevig2022} or Floquet Chern insulators\cite{Oka2019,McIver2020,Weitz2024}, suffice that sufficiently intense  photocurrents are induced by the pump. Further, we highlight that Chern number sign is directly accessible in DCD, while its magnitude is not directly connected (though possibly future spectroscopies could extend DCD to photocurrent measurements where the response can quantize in certain systems\cite{DeJuan2017} and yield this information). 

Beyond the specific Haldane nanoflakes studied here, the DCD concept is broadly applicable to a wider range of topological, magnetic, and chiral systems. It can be extended to other lattice models supporting surface modes, including quantum spin Hall and driven topological phases. It is also not restricted to electronic systems and may be adapted to photonic or phononic platforms. Moreover, it should be extendable to equivalent pump-probe time- and angle-resolved photoemission spectroscopy\cite{Ito2023,Soifer2019,Beaulieu2026,Schuler2020,Neufeld2024a,Neufeld2022d,Sidilkover2025,Gvishi2025} and transient absorption spectroscopy\cite{Wanie2024,Siegrist2019,Kobayashi2023,Lively2024,Baykusheva2022,Neufeld2023b}, and also to the linear-response regime of light-matter interactions. More generally, DCD illustrates the potential of multi-helicity control as a tool for accessing non-equilibrium topology and ultrafast chiral dynamics, and we expect it to motivate further theoretical and experimental work in broken symmetry condensed and gas phases.

\medskip

\noindent\textbf{\textit{Acknowledgments}.} The authors thank Dr. Hannes Hubener, Prof. Umberto De Giovannini, Prof. Simone Latini, and Prof. Dr. Angel Rubio, for fruitful discussions at the inception of this project several years ago. ON gratefully acknowledges support from the Young Faculty Award from the National Quantum Science and Technology program of Israel's Council of Higher Education Planning and Budgeting Committee. 

\medskip

\noindent{\textbf{\textit{Disclosure.}}} The authors declare no conflicts of interests. 

\medskip

\noindent{\textbf{\textit{Data Availability.}}} All data supporting the conclusions in this work are presented within the research paper. Additional data can be obtained from the authors upon reasonable request.

\bibliography{refs}

\end{document}


\preprint{APS/123-QED}

\title{Supplementary information for: ``Double circular dichroism high harmonic spectroscopy: An ultrafast probe for topological photocurrents"}

\maketitle

\author{Osamah Sufyan}
\affiliation{%
Department of Mathematics, Informatics, and Technology, University of Applied Sciences Koblenz, 53424 Remagen, Germany
}%

\author{Ofer Neufeld}
\email{ofern@technion.ac.il}
\affiliation{%
Technion Israel Institute of Technology, Schulich Faculty of Chemistry, Haifa, 3200003, Israel
}%

\onecolumngrid


\section{Technical details and additional results}

\noindent This SI includes all harmonic-resolved data complementary to the HHG yields presented in the main text Figures (which only presents select harmonics that show unique behaviors). 

Figures S1 and S2 present helicity-resolved yields for all harmonic orders remaining that were not displayed in main text Fig. 4. 

\FloatBarrier
\begin{figure*}[h]
    \centering
    \includegraphics[width=1\textwidth]{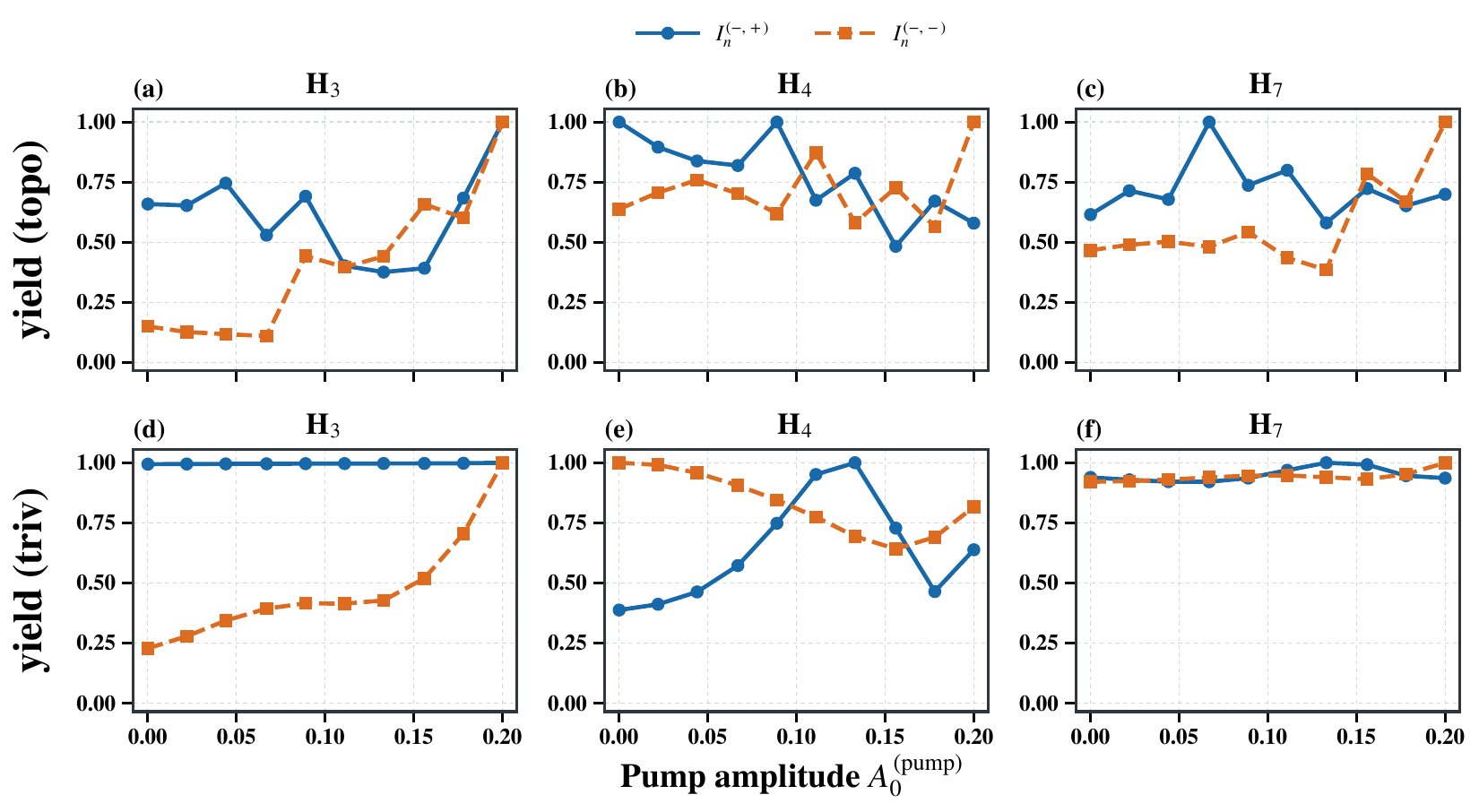}

\caption{
        Same as Fig. 4 in the main text but for remaining harmonic orders (3,4,7). Each harmonic order is indicated in plot.}
    \label{fig:hhg_yields_topo_trivial_remaining}
\end{figure*}

\begin{figure*}[h]
    \centering
    \includegraphics[width=1\textwidth]{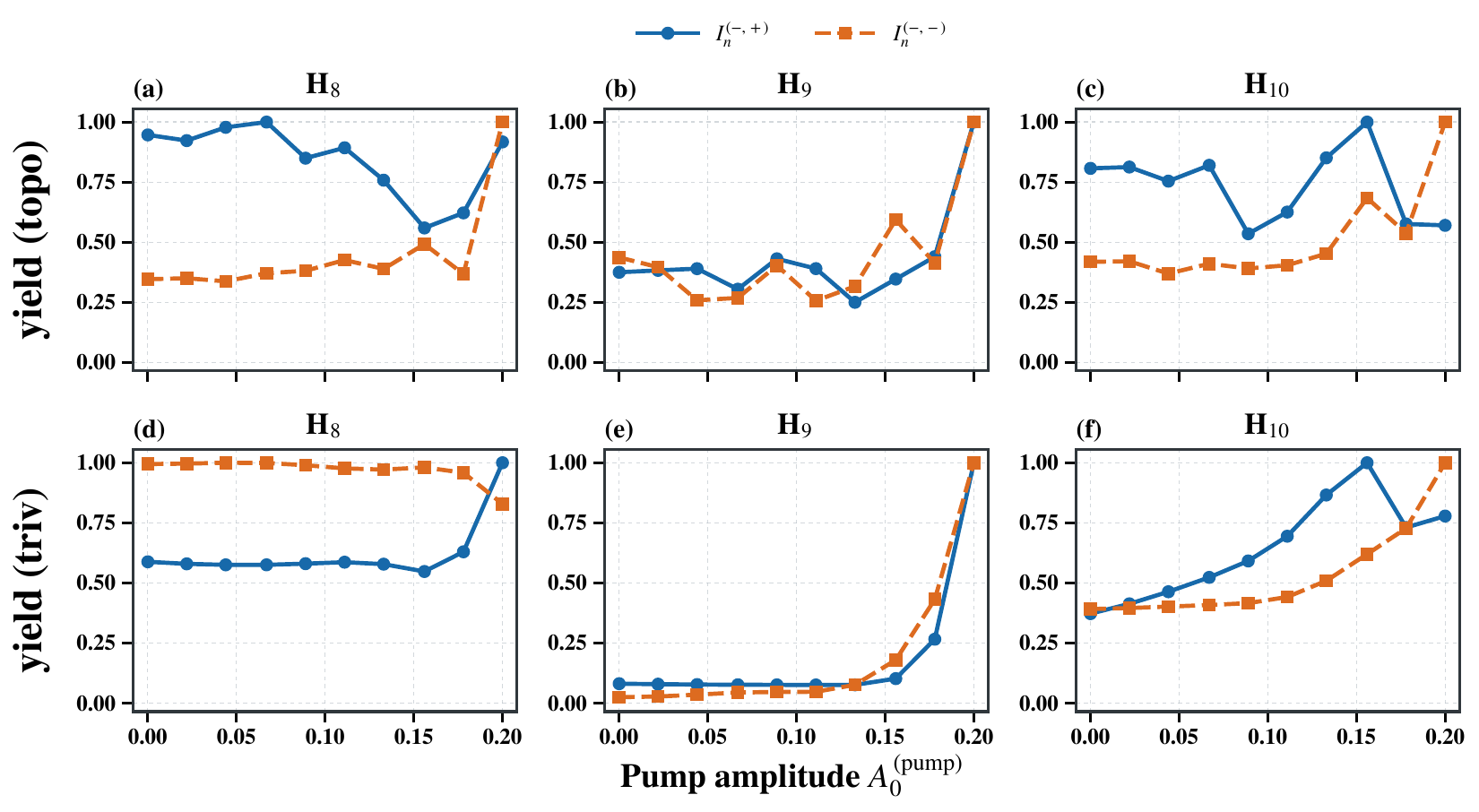}

\caption{
        Same as Fig. S1 but for remaining harmonic orders (8,9,10). Each harmonic order is indicated in plot.}
    \label{fig:hhg_yields_topo_trivial_remaining}
\end{figure*}
\FloatBarrier

Figure S3 presents the helicity-resolved CD and resulting DCD for all harmonic orders missing in the main text Fig. 5. 

\FloatBarrier
\begin{figure*}[t]
    \centering
    \includegraphics[width=\textwidth]{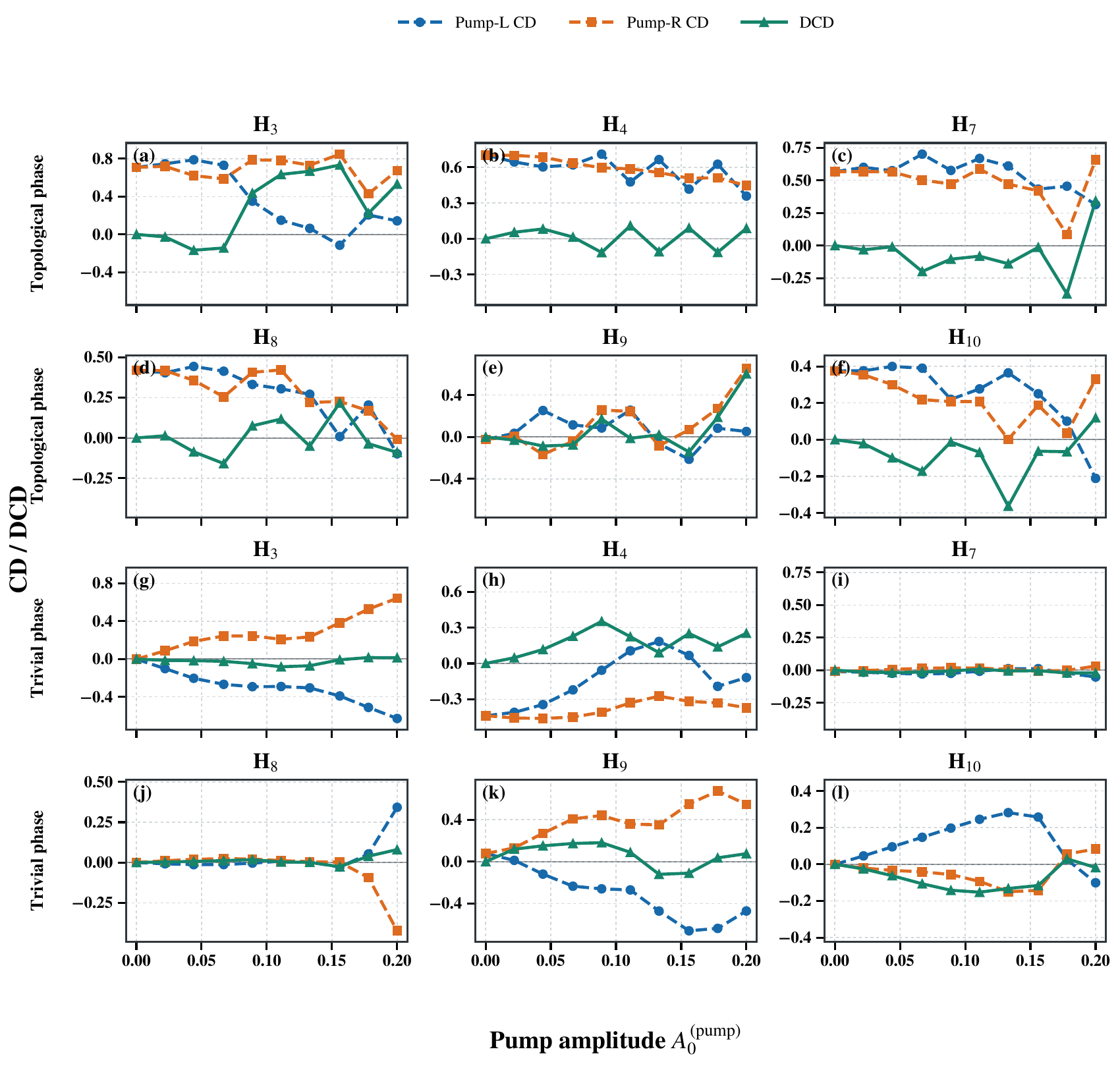}

    \caption{
        Same as Fig. 5 in the main text but for the remaining harmonic orders. The green curves use the bounded DCD definition given in the main text.
    }

    \label{fig:cd_A0_topo_trivial_remaining}
\end{figure*}
\FloatBarrier

Figure S4 presents the CD and DCD responses for select harmonics orders (as discussed in the main text) but vs the system size, showing that the DCD is stable with the flake size (unlike the generic CD response). 

\FloatBarrier
\begin{figure*}[h]
\centering

    \includegraphics[width=0.8\textwidth]{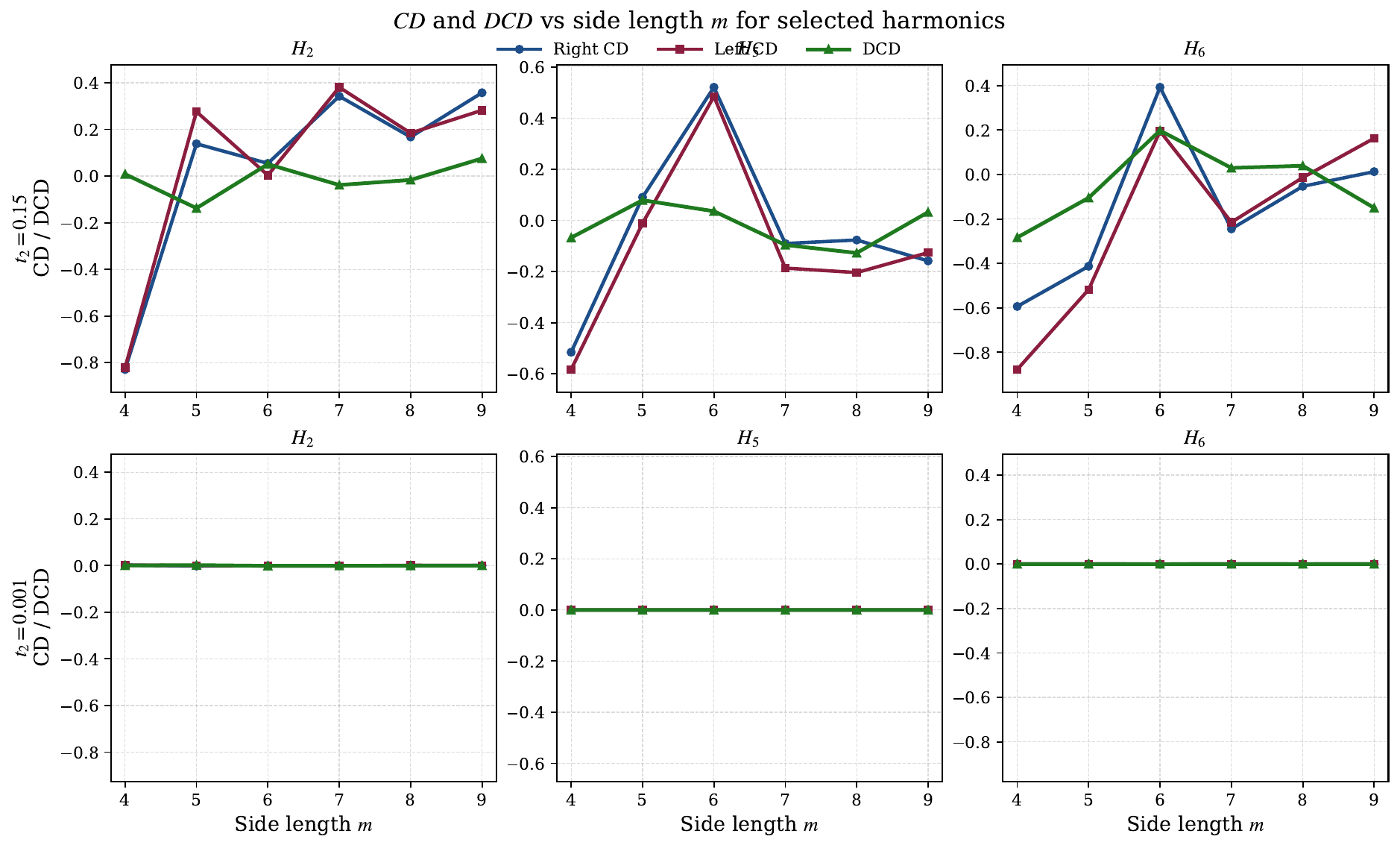}

\caption{HHG CD and DCD for select harmonic orders as in the main text Fig. 5, but vs. the system size ($m$ - the number of sites on one Haldane flake edge). We show that the CD oscillates strongly with system size, but the DCD is relatively stable.}

\label{fig:currents_combined}
\end{figure*}
\FloatBarrier

Figure S5 presents the CD and DCD across the topological phase transition in similar conditions to Fig. 8 in the main text, but for the remaining harmonic orders. The response is seen to be very similar across orders, with a sharp change in the strong oscillations in CD and DCD at the phase transition. Generally the CD and DCD do not vanish in the trivial phase, but simply become roughly parameter-independent, or linearly-dependent, eliminating oscillations (as discussed in the main text). We largely see that DCD exhibits less sharp oscillations in the topological phase and is more stable. 

\FloatBarrier
\begin{figure*}[h]
\centering

    \includegraphics[width=0.8\textwidth]{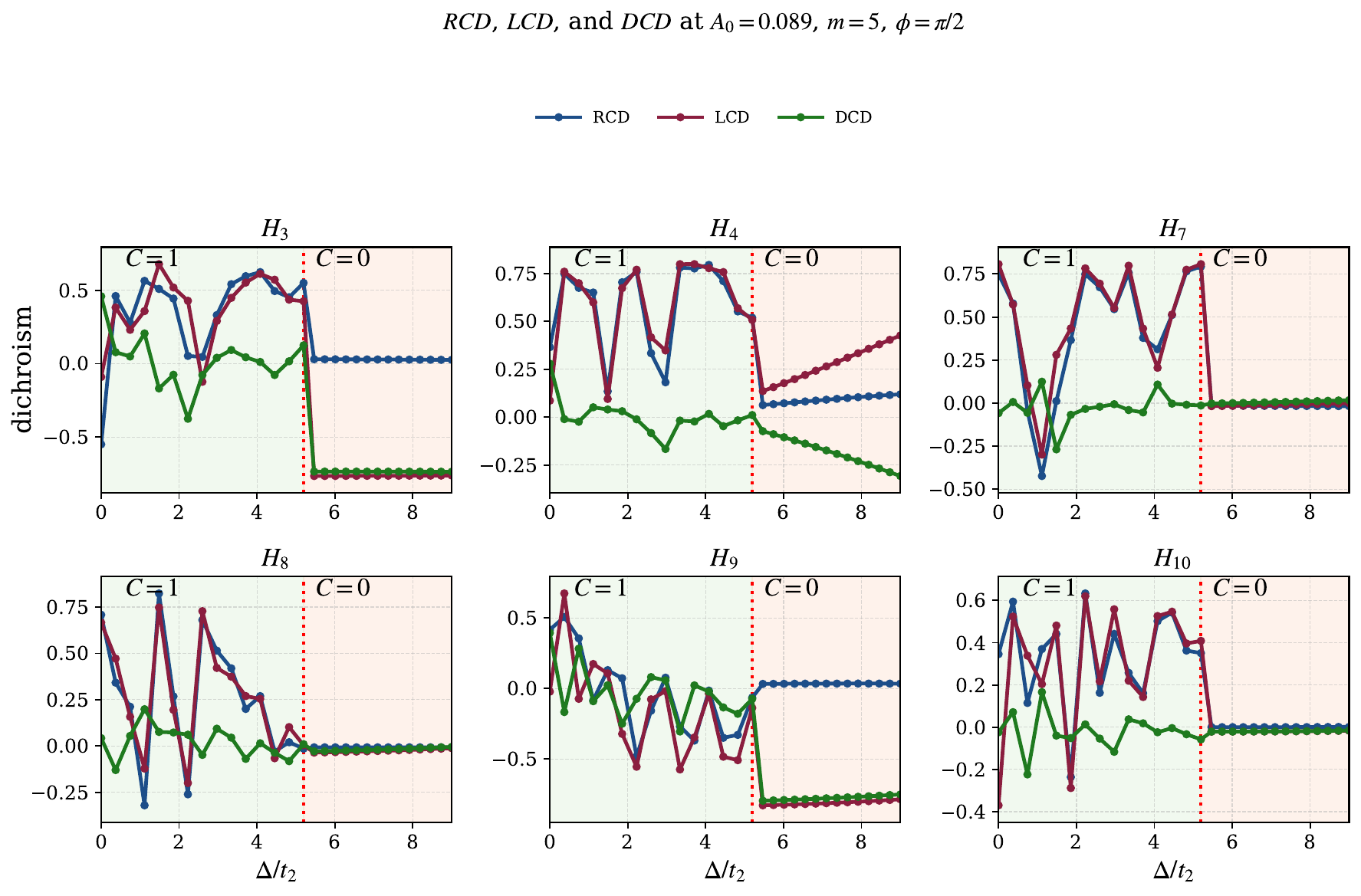}

\caption{HHG CD and DCD across the phase transition as shown in Fig. 8 in the main text, but for the remaining harmonic orders.}

\label{fig:currents_combined}
\end{figure*}
\FloatBarrier

Figures S6 and S7 together show the HHG CD and DCD comparison between the boundary-truncated and coherent full-flake calculations displayed in main-text Figure 7, but for the remaining harmonics not displayed in the main text.

\FloatBarrier
\begin{figure*}[h]
    \centering
    \includegraphics[width=\textwidth]{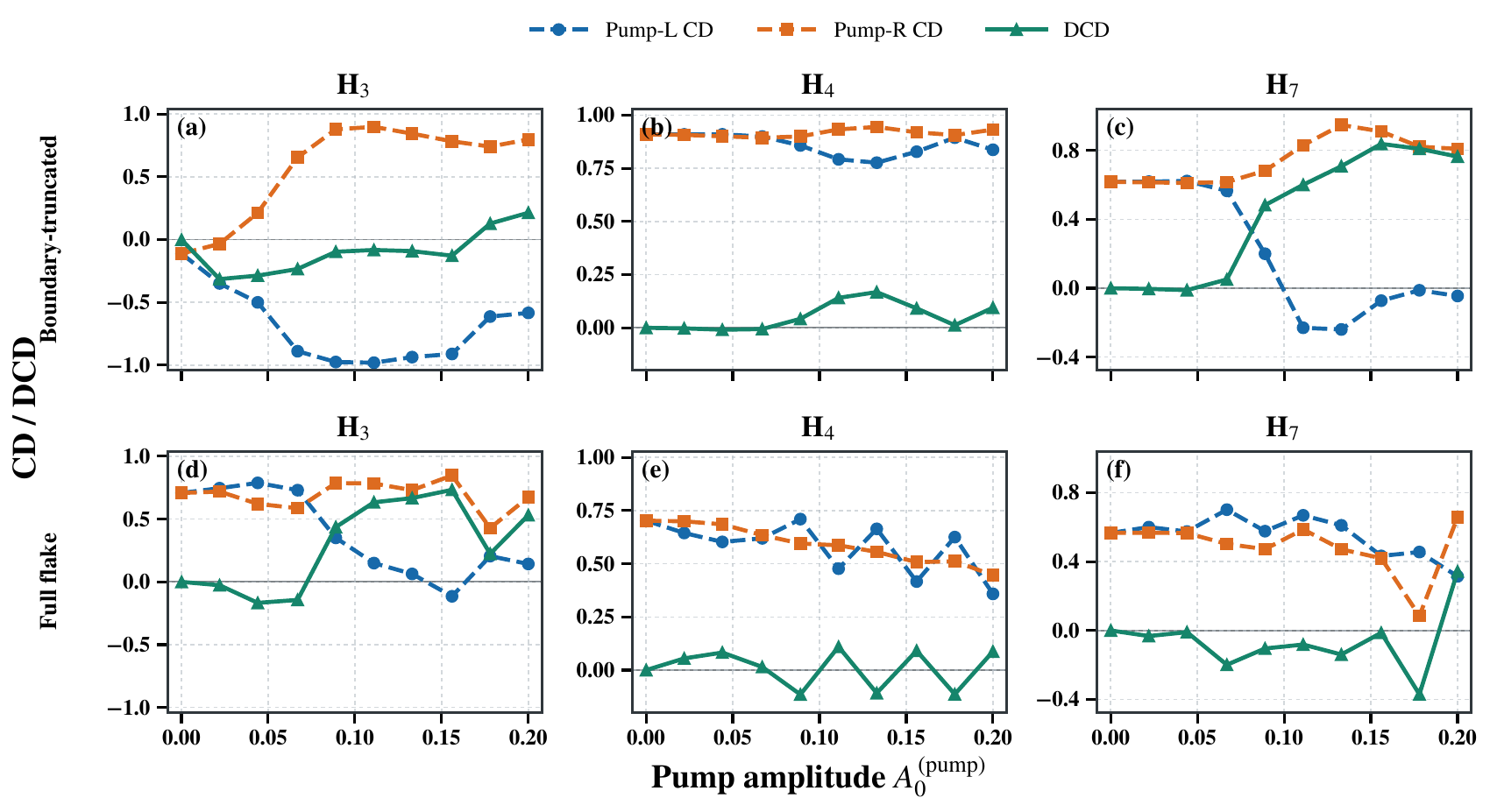}

    \caption{
        Same as Fig. 7 in the main text, but for harmonic orders 3, 4, and 7. The upper and lower rows show the boundary-truncated and coherent full-flake calculations, respectively; the green curves use the bounded DCD definition.
    }
    \label{fig:no_edge_comparison_part1}
\end{figure*}

\begin{figure*}[h]
    \centering
    \includegraphics[width=\textwidth]{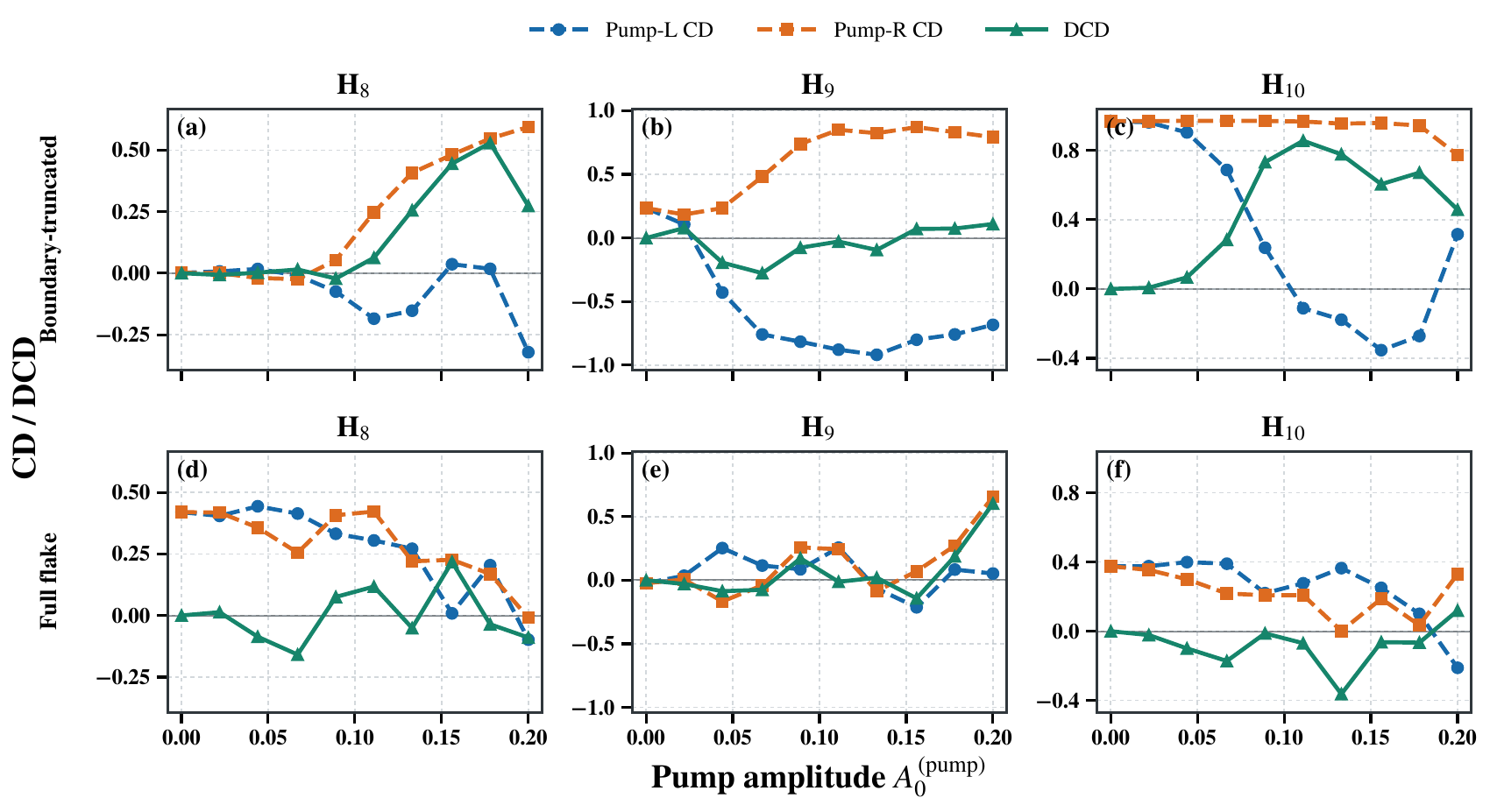}

    \caption{
        Same as Fig. 7 in the main text, but for harmonic orders 8, 9, and 10. The upper and lower rows show the boundary-truncated and coherent full-flake calculations, respectively; the green curves use the bounded DCD definition.
    }
    \label{fig:no_edge_comparison_part2}
\end{figure*}
\FloatBarrier

Figure S8 presents the delay-dependent CD and DCD for all studied harmonic orders not displayed in the corresponding main-text figure.

\FloatBarrier
\begin{figure*}[h]
    \centering
    \includegraphics[width=\textwidth]{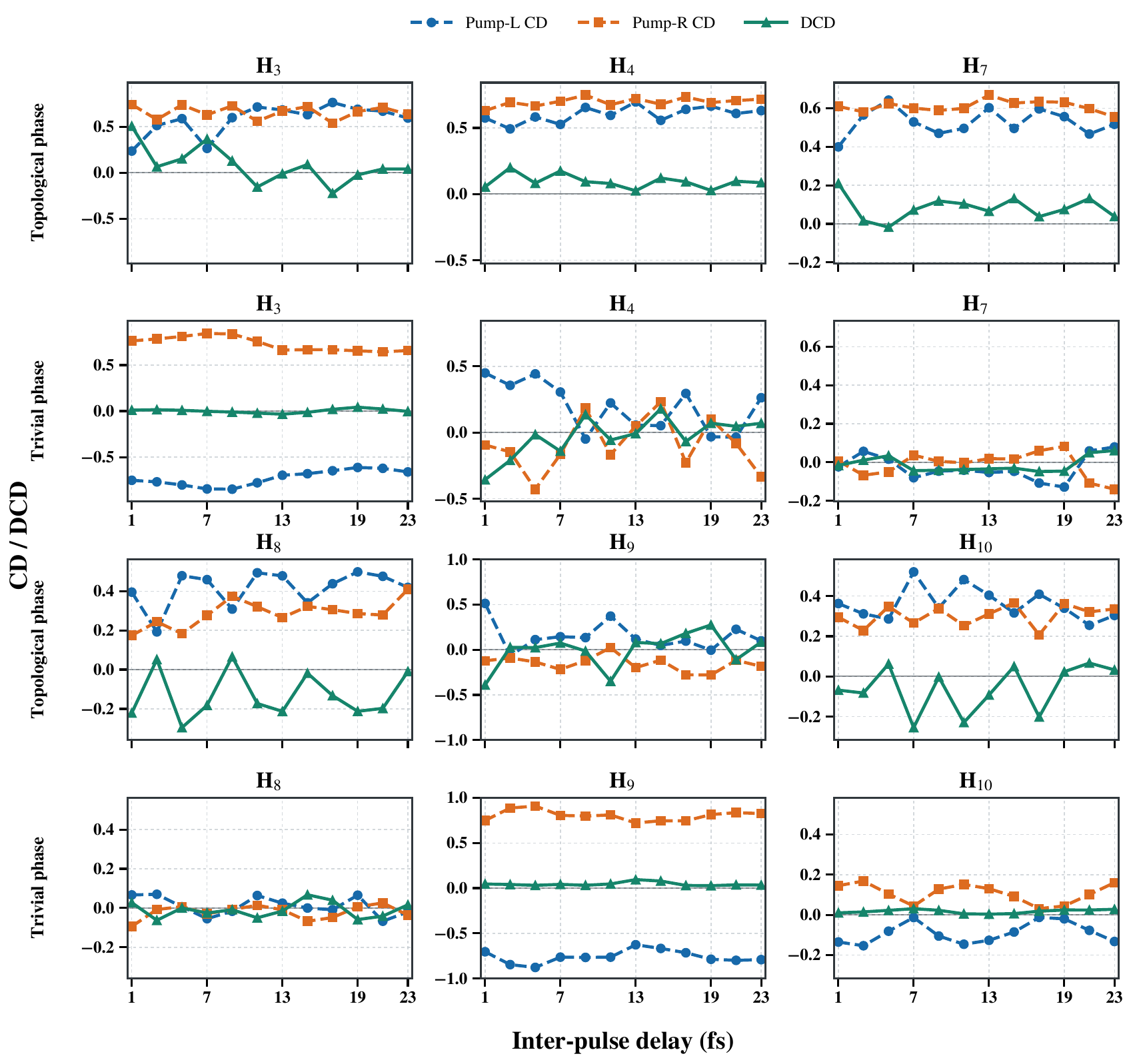}
    \caption{Delay dependence of the pump-R CD, pump-L CD, and DCD for harmonic orders 3, 4, 7, 8, 9, and 10 in the topological and trivial phases. The parameters and conventions are the same as in the corresponding main-text figure. The horizontal axis gives the field-free interval between the pump and probe pulses. Harmonic order 1 is omitted because it is not studied in this work.}
    \label{fig:delay_cd_dcd_remaining}
\end{figure*}
\FloatBarrier

Figure S9 presents the time-step convergence of the representative topological HHG spectrum shown in the inset of main-text Fig. 1.
\FloatBarrier
\begin{figure*}[h]
    \centering
    \includegraphics[width=0.52\textwidth5]{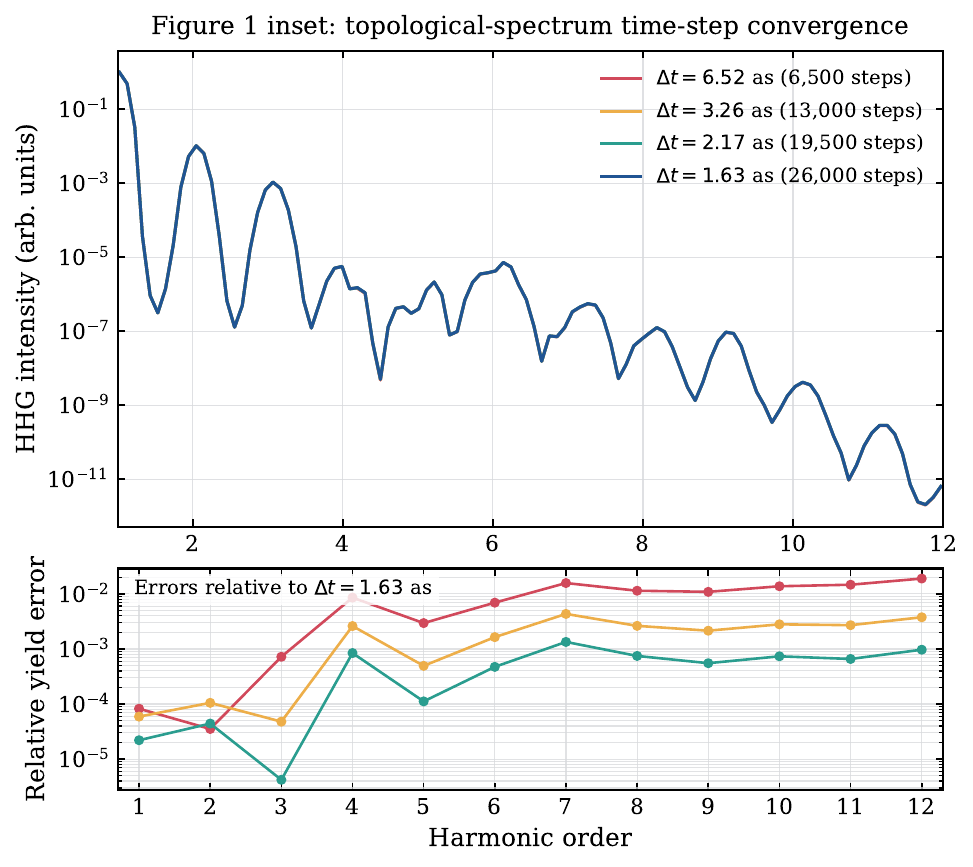}
    \caption{Time-step convergence of HHG spectra in similar conditions to main-text Fig. 1. The upper panel compares the spectra on a common intensity scale, showing virtually fully converged spectra. The lower panel shows the relative error of the integrated H1--H12 yields with respect to the finest time step. For $\Delta t=3.26$ as used for the original spectrum the median relative error over H1--H12 is $0.24\%$, demonstrating tight numerical convergence.}
    \label{fig:fig1_timestep_convergence}
\end{figure*}
\FloatBarrier

Figure S10 presents DCD exploration in graphene, i.e. a system respecting time-reversal symmetry, by setting $t_2=\Delta=0$ in our simulations. This shows that the DCD respects the appropriate limits and vanishes identically in non-symmetric systems, though the CD from the pumped systems does not necessarily vanish since the pump can induce a photocurrent in the bulk.

\begin{figure*}[t]
    \centering
    \includegraphics[width=0.7\textwidth]{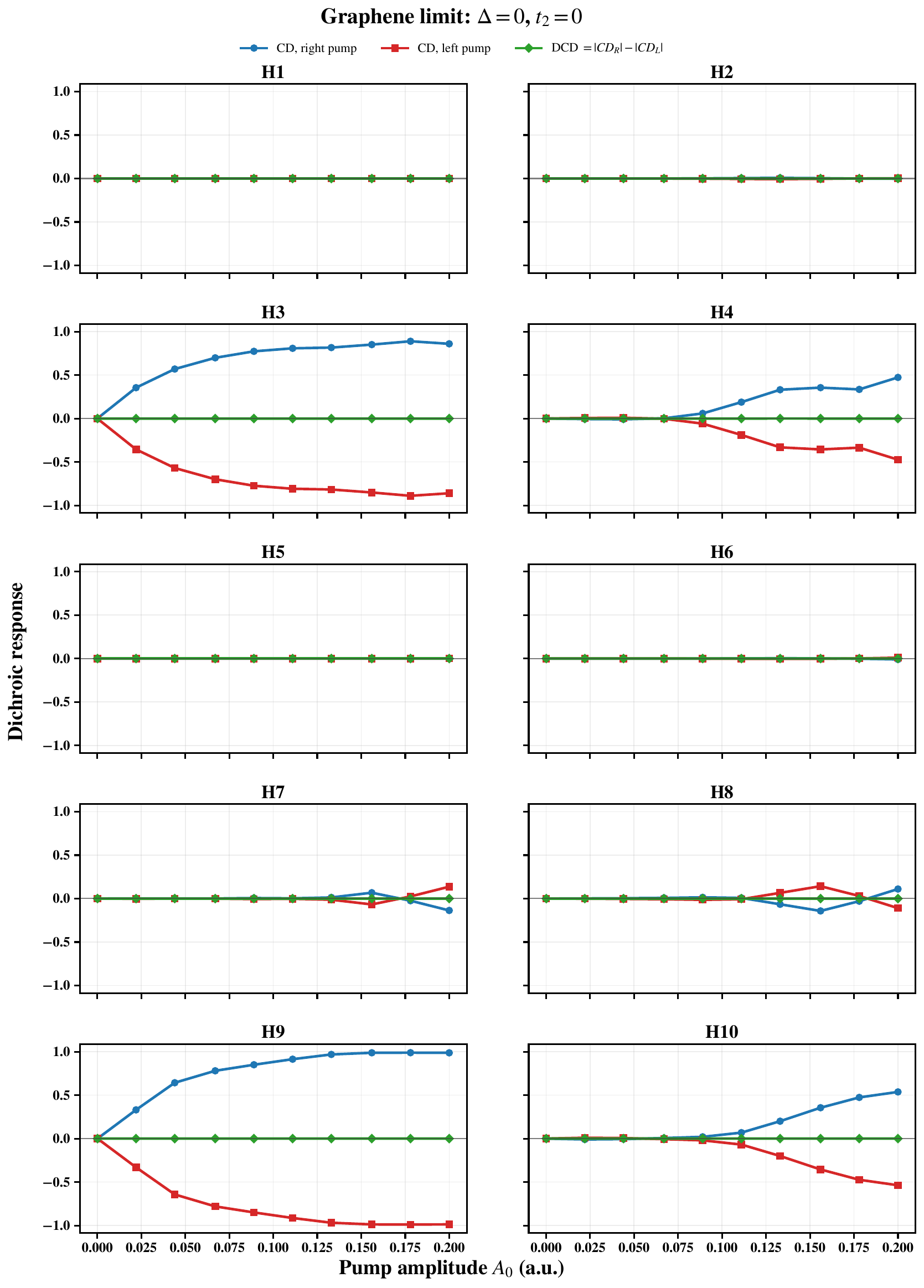}
    \caption{DCD from a model graphene ($\Delta=t_2=0$) for H1--H10, showing vanishing DCD despite non-vanishing CD from the pumped system. This cancellation is a result of the DCD construction, which vanishes in systems with time-reversal symmetry.}
    \label{fig:graphene_absdiff_control}
\end{figure*}

\clearpage
\section{Bulk--edge interference and edge-state localization analysis in DCD}

Across the text there is a main question concerning the role of interference of bulk/edge components in emerging HHG and DCD signals. We now directly explore this point with several approaches. First, Fig. S11 tests the robustness of the boundary-truncated DCD against defining the removed boundary as one, two, or three outer atomic layers (the main text displays only the single-layer removed data). The reason for performing this analysis is that the localized protected edge modes in principle to not localized purely over the first atomic layer at the flake boundary, but also have some small exponentially-decaying contributions towards the bulk in the additional 2nd and 3rd layers of atoms. We wish now to resolve if those contributions paly a key role in the surface-contributed DCD. The results in Fig. S11 show that, largely, the DCD is quite insensitive across most harmonic orders to additional truncation beyond the first layer. This implies that indeed the most dominant surface-related topological photocurrent response arises from just the first layer, as analyzed in the main text. It suggests that DCD can indeed be a good measure of edge-state localized physics since the inner layers coupling do not greatly impact it, but simply follow bulk-like physics.

\FloatBarrier
\begin{figure*}[h]
    \centering
    \includegraphics[width=\textwidth]{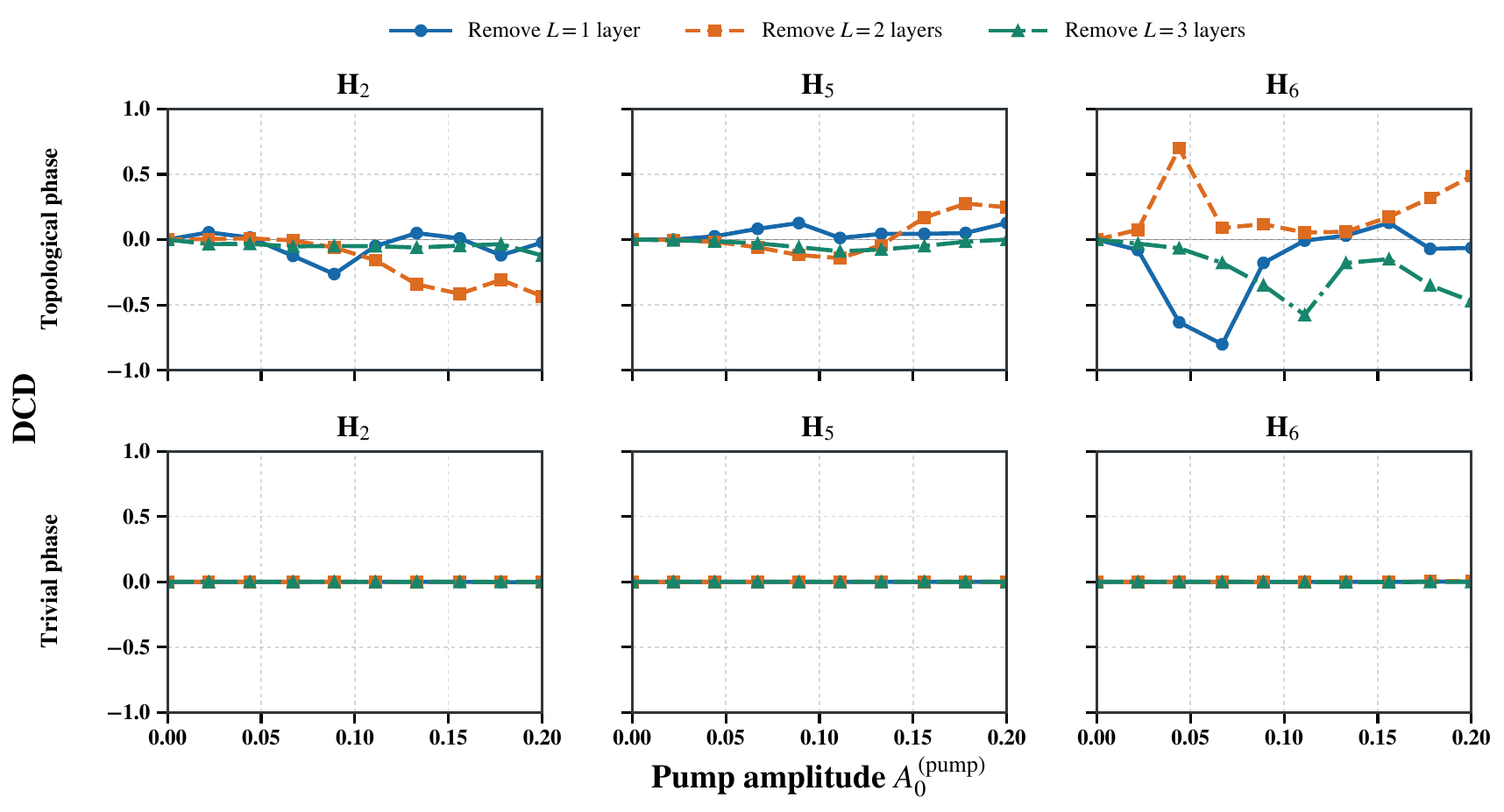}
    \caption{Boundary-robustness of the bounded DCD for H2, H5, and H6. The boundary-truncated calculation is repeated after removing one (as in main text), two, or three outer atomic layers.}
    \label{fig:boundary_layer_robustness}
\end{figure*}
\FloatBarrier

Moving further, we wish to test the role of bulk/edge interference in the emitted DCD signal. The full-flake HHG spectrum is calculated from the total time-dependent current, and therefore contains coherent interference of such terms by construction. To expose this interference explicitly, we introduce an operational boundary mask and explicitly dissect current matrix elements into bulk--bulk ($BB$), edge--edge ($EE$), and bulk--edge ($BE$) sectors. For harmonic order $n$, the complex two-component spectral amplitude is
\begin{equation}
    \mathbf{A}_{n}=\mathbf{A}_{BB,n}+\mathbf{A}_{EE,n}+\mathbf{A}_{BE,n},
    \label{eq:sector_amplitude}
\end{equation}
where each vector contains the Fourier-transformed $x$ and $y$ acceleration amplitudes integrated over the corresponding harmonic window. The full coherently-summed harmonic intensity is consequently
\begin{align}
    I_n^{\mathrm{coh}}
    ={}& \sum_{q\in\{BB,EE,BE\}}\left\|\mathbf{A}_{q,n}\right\|^2 \nonumber\\
    &+\sum_{q<r}2\,\mathrm{Re}\!\left[\mathbf{A}_{q,n}\!\cdot\!\mathbf{A}_{r,n}^{*}\right].
    \label{eq:coherent_intensity}
\end{align}

In contrast, by omitting amplitude-level interference we obtain a pure self-term reference (which would be akin to a pure incoherent summation of the signal from separated bulk/edge regions):
\begin{equation}
    I_n^{\mathrm{self}}=\sum_{q\in\{BB,EE,BE\}}\left\|\mathbf{A}_{q,n}\right\|^2.
    \label{eq:self_intensity}
\end{equation}
We can therefore evaluate the absolute ratio between the interference terms compared to the incoherent summation by defining:
\begin{equation}
    \mathcal{R}_n=
    \frac{\displaystyle\sum_{q<r}\left|2\,\mathrm{Re}\!\left[\mathbf{A}_{q,n}\!\cdot\!\mathbf{A}_{r,n}^{*}\right]\right|}
    {\displaystyle\sum_q\left\|\mathbf{A}_{q,n}\right\|^2}.
    \label{eq:interference_ratio}
\end{equation}
Mathematically, $\mathcal{R}_n\sim1$ compares the absolute cross-term weight to the summed self terms --- if it is large, that means interference of bulk/edge emission plays a major role in pump-probe HHG from topological systems, while if it is small it implies that the bulk/edge signals are well separated. Circular dichroism can be further evaluated separately for each pump helicity $s\in\{R,L\}$, such the similar quantities can be defined for CD and DCD:
\begin{equation}
    \mathrm{CD}_{s,n}=\frac{I_{sR,n}-I_{sL,n}}{I_{sR,n}+I_{sL,n}},
    \qquad
    \mathrm{DCD}_n=|\mathrm{CD}_{R,n}|-|\mathrm{CD}_{L,n}|.
    \label{eq:interference_dcd}
\end{equation}
The sector amplitudes in Eq.~\ref{eq:sector_amplitude} reconstructs the directly evaluated total current. For the manuscript-matched H2, H5, and H6, the reconstructed and directly evaluated bounded DCD agree to a maximum absolute difference of $5.1\times10^{-12}$, providing a stringent numerical check of the decomposition. Note that the decomposition is used here as a numerical trick only to quantify coherent interference and should not interpreted as a unique experimental separation of pure bulk and edge emitters. However, if interference is small, it implies potentially such separation could be achieved by a more elaborate scheme. Figure S12 quantifies the total magnitude of the interference terms relative to the self-term intensity, while Figure S13 shows how the bounded DCD changes as the interference terms are added successively. Crucially, this analysis shows that DCD is relatively weakly affected by the interference (Fig. S13), unlike the pure direct HHG yields (Fig. S12). This supports the notion that HHG-DCD can be a more robust approach for bulk/edge separation discussed in the main text, and implies that a main contribution from the cross terms cancels out in the DCD subtractions of HHG yields for several configurations.

\FloatBarrier
\begin{figure*}[h]
    \centering
    \includegraphics[width=\textwidth]{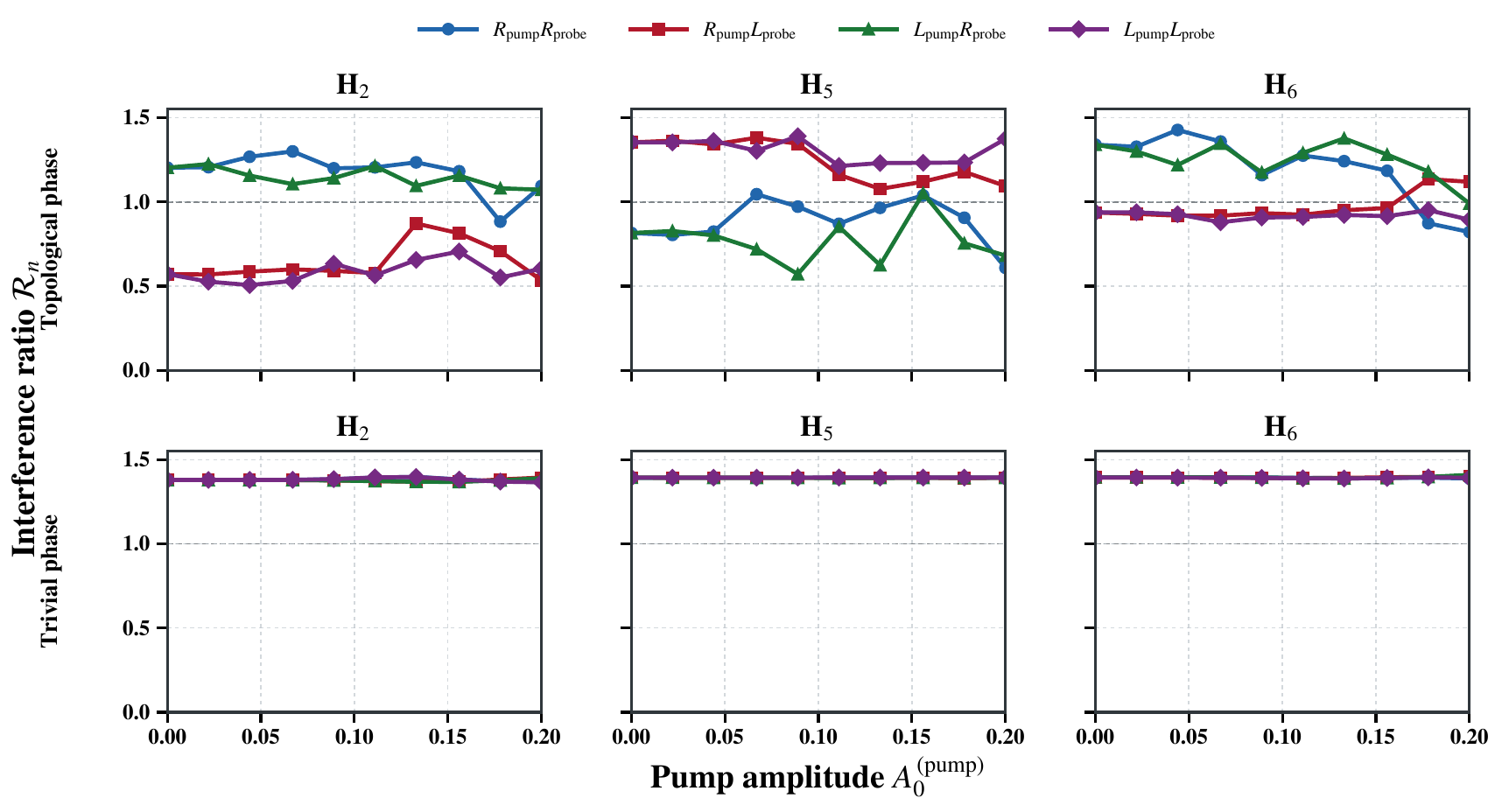}
    \caption{Absolute interference ratio $\mathcal{R}_n$, Eq.~\ref{eq:interference_ratio}, for H2, H5, and H6 and all four pump--probe helicity configurations. The dashed horizontal line marks interference equal in magnitude to the self-term sum. Across the sampled pump amplitudes and helicities, the median (maximum) ratios are 1.042 (1.427) in the topological calculation and 1.392 (1.409) in the trivial calculation. The cross terms are therefore not perturbatively small in either phase and clearly affect HHG.}
    \label{fig:interference_ratio}
\end{figure*}

\begin{figure*}[h]
    \centering
    \includegraphics[width=\textwidth]{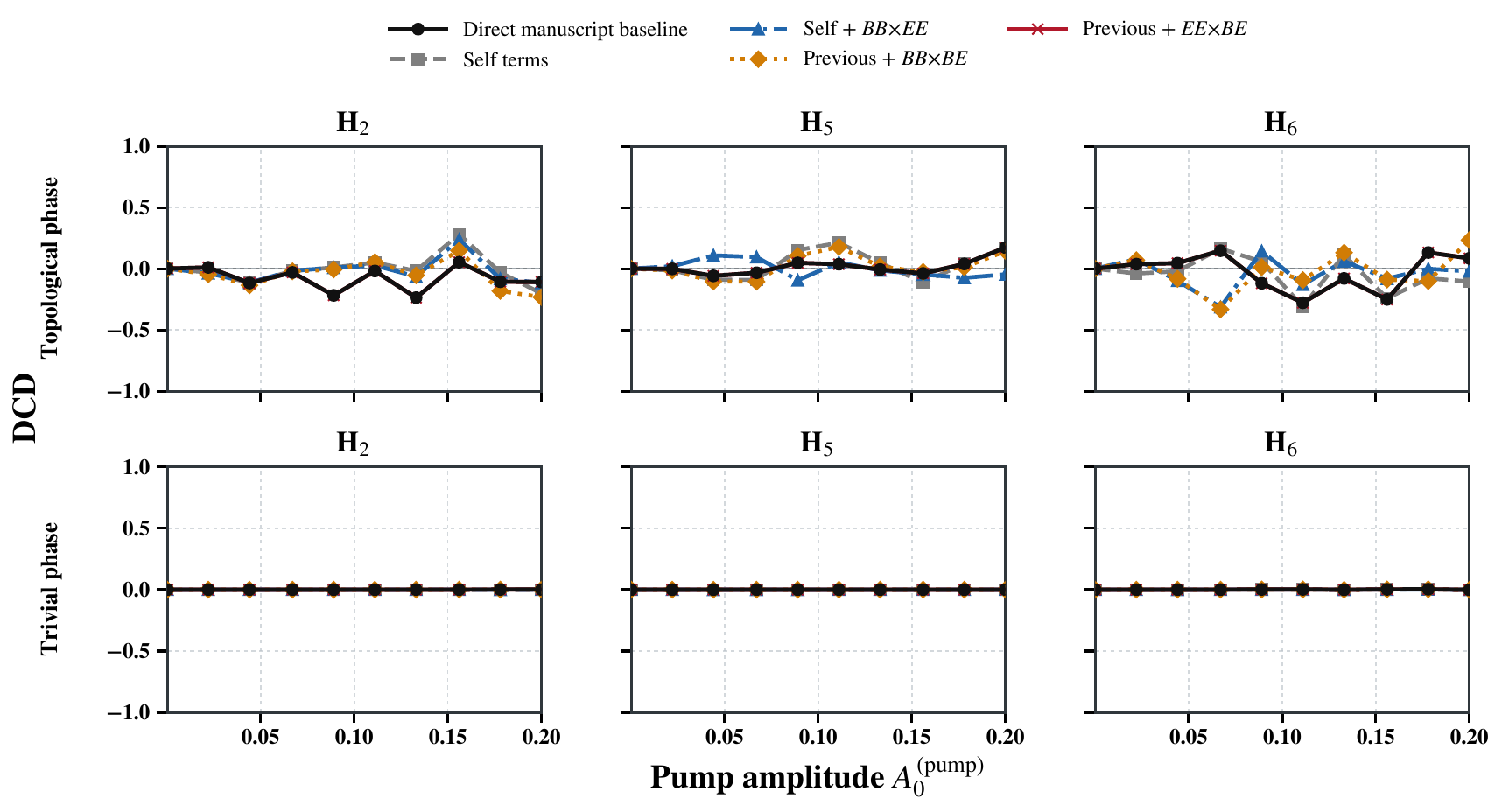}
    \caption{Bounded DCD reconstructed by adding the three amplitude-level interference terms successively to the self-term reference. The black curves are obtained directly from the total manuscript current, while the final red curves contain all self and cross terms from Eq.~\ref{eq:coherent_intensity}. Their numerical agreement verifies the coherent reconstruction.}
    \label{fig:interference_reconstruction}
\end{figure*}